\documentclass{aa}

\usepackage{graphicx}
\usepackage{newtxtext}
\usepackage{newtxmath}

\usepackage{multirow}
\usepackage{longtable}
\usepackage[left,modulo,switch]{lineno} 
\usepackage{array}
\usepackage{lipsum}
 \usepackage{ulem} 
\usepackage[usenames,dvipsnames,svgnames,table]{xcolor} 

\usepackage{hyperref}
\hypersetup{
  colorlinks=true,
  citecolor=blue,
  linkbordercolor={1 0 0},
  linkcolor=blue,
  urlcolor=blue,
  breaklinks=true
}

\newcommand{\ie}{i.e.}

\newcommand{\rrlyr}{RR~Lyrae\ }
\newcommand{\rrlyrn}{RR~Lyrae}
\newcommand{\rrly}{RR~Lyr\ }
\newcommand{\rrlyn}{RR~Lyr}
\newcommand{\ha}{H$\alpha$\ }

\newcommand{\he}{He\,I\ } 

\newcommand{\bla}{Blazhko\ }
\newcommand{\kms}{\;km\,s$^{-1}$\ }
\newcommand{\kmsn}{\;km\,s$^{-1}$}

\newcommand{\itv}[1]{\textit{#1}}

\renewcommand{\aap}{A\textit{\&}A} 

\begin{document}

\title{Dynamical structure of the pulsating atmosphere of \rrlyr}

\subtitle{I. A typical pulsation cycle}

\author{
  D. Gillet\inst{1}
  \and B. Mauclaire\inst{2}
  \and T.~Lemoult\inst{3}
  \and Ph. Mathias\inst{4}
  \and J-S. Devaux\inst{5}
  \and Th. de France\inst{6}
  \and T.~Garrel\inst{7}
  \and the GRRR Collaboration\inst{8}\thanks{
  \textit{The \textbf{G}roupe de \textbf{R}echerche sur \textbf{RR} Lyrae} (GRRR) is an association of professionals and amateur astronomers leading high-resolution spectroscopic and photometric monitoring of complex phenomena such as the \rrlyr Blazhko effect.}
}

\institute{
  Observatoire de Haute-Provence -- CNRS/PYTHEAS/Université d'Aix-Marseille, 04870 Saint-Michel l'Observatoire, France \\ \email{denis.gillet@osupytheas.fr}
  \and Observatoire du Val de l'Arc, route de Peynier, 13530 Trets, France \\ \email{bma.ova@gmail.com}
  \and Observatoire de Chelles, 23 avenue Hénin, 77500 Chelles, France
  \and IRAP, Universit\'e de Toulouse, CNRS, UPS, CNES, 57 avenue d'Azereix, 65000, Tarbes, France
  \and Observatoire OAV, 13 rue du Moulin, 34290 Alignan-du-Vent, France
  \and Observatoire des Tourterelles, 5 impasse Tourterelles, 34140 Mèze, France 
  \and Observatoire de Fontcaude, 19 avenue du hameau du Golf, 34990 Juvignac, France 
  \and Observatoire de Haute-Provence, 04870 Saint-Michel l'Observatoire, France 
  \thanks{Based in part on observations made at Observatoire de Chelles, 77500 Chelles, France}
}

\date{Received 16 July 2018 / Accepted 16 January 2019.}

\abstract
{RRab stars are large amplitude pulsating stars in which the pulsation wave is a progressive wave. 
Consequently, strong shocks, stratification effects, and phase lag may exist between the variations associated with line profiles formed in 
different parts of the atmosphere, including the shock wake. 
The pulsation is associated with a large extension of the expanding atmosphere, and strong infalling motions are expected.}
{The objective of this study is to provide a general overview of the dynamical structure of the atmosphere occurring over a typical pulsation cycle.}
{We report new high-resolution observations with suitable time resolution of \ha and sodium lines in the brightest RR\,Lyrae star of the sky: RR\,Lyr (HD 182989). 
A detailed analysis of line profile variations over the whole pulsation cycle is performed to understand the dynamical structure of the atmosphere. 
}
{The main shock wave appears when it exits from the photosphere at $\varphi\simeq0.89$, \ie, when the main \ha emission is observed. 
Whereas the acceleration phase of the shock is not observed, a significant deceleration of the shock front velocity is clearly present. 
The radiative stage of the shock wave is short: $4\%$ of the pulsation period ($0.892<\varphi<0.929$). 
A Mach number $M>10$ is required to get such a radiative shock. 
The sodium layer reaches its maximum expansion well before that of \ha ($\Delta\varphi=0.135$). 
Thus, a rarefaction wave is induced between the \ha and sodium layers.
A strong atmospheric compression occurring around $\varphi=0.36$, which produces the third \ha emission, takes place in the highest part of the atmosphere. 
The region located lower in the atmosphere where  the sodium line is formed is not involved.
The amplification of gas turbulence seems mainly due to strong shock waves propagating in the atmosphere rather than to the global compression of the atmosphere caused by the pulsation. It has not yet been clearly established whether the microturbulence velocity increases or decreases with height in the atmosphere.
Furthermore, it seems very probable that an interstellar component is visible within the sodium profile.
}
{}



\maketitle

\section{Introduction}


The kinematics and dynamics of the outer layers of pulsating stars, {a fortiori} those undergoing Blazhko effect \citep{bla1907}, is usually not known in detail. Among variable stars, \rrly is the brightest \rrlyr star of the sky with $7\la\textrm{V}\la8$. During its pulsation cycle, its photospheric radius relative variation $\Delta R_\star/R_\star=17\%$ \citep{fokin97}, while is spectral type evolves  from A7III to F8III \citep{gillet88}.
\rrly variability was discovered by the Scottish astronomer Williamina Fleming at Harvard in 1901 \citep{pick1901}.  
The pulsation cycle occurs on approximately 0.5668\,d or, equivalently, 13.6\,h.
Furthermore, the light curve also presents light modulations with a variable period around 39\,d. 
These amplitude and phase modulations are  known as the Blazhko effect. 
Its physical origin has remained a mystery up to the present day, but recently several interesting ideas were proposed to explain the Blazhko phenomenon \citep{smolec13, kovacs16}.
All these phenomena are the rendering of atmospheric movements that are not  simple outward and inward movements that may show light curves (see RR41 run of \citealt{fokin97}).


\medskip

In order to study the atmospheric dynamics, the whole pulsation cycle has to be recorded in detail.
The first detailed spectroscopic study of \rrly was done by \citet{preston65} during a Blazhko cycle, but this investigation was 
limited to the pulsation phases of rising light within the 13.6\,h cycle. 
\cite{chadid96} carried on spectroscopic observation and showed metallic line doubling in \rrlyn. 
They suggested conducting some extra observations with a larger telescope to get a higher S/N and time resolution.
\cite{preston11} and \cite{chadid13} performed extra spectroscopic observations. 
They studied dynamics of various \rrlyr stars, but did not focus on \rrly itself.
\cite{fossati2014} carried out a spectroscopic abundance analysis of RR Lyr over a pulsation cycle. 
This analysis was centered on the so-called “quiet phase” around the maximum radius. 
In particular, they show that the microturbulent velocity $V_{mic}$ close to the photospheric level is not significantly affected by the 
pulsation and remains almost constant during the pulsation cycle while there is an increase in $V_{mic}$ around the bump phase and again 
on the rising branch. 
During these two phases, strong shocks propagate rapidly through the atmosphere.
\cite{fossati2014} confirm the presence of a strong peak of the turbulent velocity around the pulsation phase 0.9.
Nevertheless, none of these historical surveys gave a description of the whole atmospheric dynamic structure during a pulsation cycle.
A recent paper \citep{gillet2017a} presents an additional spectroscopic survey dedicated to the ``third apparition.''
Since they saw an interesting behavior in the sodium doublet region, we decided to conduct a more intensive run in order to get a 
more detailed view of \rrly dynamics.


\medskip

In order to find out the origins of the atmospheric layer dynamics, it is necessary to compare observations with relevant atmospheric models, 
but this is only possible if an observational survey with a high temporal resolution during the whole pulsation cycle is available.
There are many models that describe photosphere evolution. 
Unfortunately, only a small number of them integrate a description of the atmosphere and the photosphere together with an adapted treatment of shock waves.
\cite{1999A&A...351..103F} developed a nonlinear convective model of pulsating stars with shock waves based on the Vienna model. 
Computed light curve morphology is in good agreement with observed RR\,Lyrae light curves, including the well known \rrlyr phase discrepancy problem \citep{1985ApJ...299..723S}. 
Recently, \cite{2013ApJ...771..113G} implemented a two-dimensional (2D) radiation hydrodynamic code that simulates the interaction of radial 
pulsation and convection. 
The code is able to match the observed light curve shape near the red edge of the RR\,Lyrae instability strip the in HR diagram. 
However, neither of these  models is able to compute a spectral line profile.

While convection is a tricky phenomenon to model, only a few pulsation models are able to calculate line profiles.
\cite{hill72} built a nonlinear hydrodynamic model of atmosphere with shock waves which leads to a successful comparison of light curves, 
absorption line velocities, emission strengths, and shock wave dissipation for the RR\,Lyrae star X\,Ari. 
\cite{1975ApJ...195..441S} described a nonlinear hydrodynamic model atmosphere with shock waves taking into account dissipation. 
It is able to compute light curves, growth rates, layer temperatures, and velocities during the pulsation cycle.
More recently, \cite{2013arXiv1310.0535S} developed an adiabatic three-dimensional (3D) pulsation model for RR\,Lyrae stars that he applied to RR\,Lyr itself. 
It considers the outer boundary condition and the interaction between pulsation and turbulence, but this model does not include radiative effects such as 
radiation and ionization. 
In these conditions, turbulent mixing and the dynamics of an extended expanding atmosphere produce shocks with extreme intensities. 
However, such large intensities have not been observed yet.

A work dedicated to RR\,Lyr was done by \cite{fokin97}. 
They built a series of nonlinear, nonadiabatic models including atmosphere and shock wave propagation. 
In particular, they used a purely radiative approach because convection is thought to strongly affect the helium ionization zone of RR\,Lyr according to its 
effective temperature \citep{XCD98}. 
The resulting \ion{Fe}{II} and \ion{Ba}{II} line profiles have been  confirmed by high-resolution observations such as line doubling phenomena according to 
the \cite{schwa52} mechanism. 
Lines profiles are computed at any pulsation phase according to predicted physical parameters such as temperature, density, or turbulent velocity occurring within the line-forming region. 
This seems to be correct when lines are formed over large areas of the atmosphere (absorption lines). 
On the other hand, when lines are produced in the radiative wake of the shock, such as \ha emission, line profiles are only approximate because 
non-LTE effects are not fully taken into account.
In addition, new high-quality spectral observations around the relevant pulsation phases may involve improvements in this kind of model.

There are currently too few pulsation atmospheric models that deal with both convection and shock waves that are able to compute synthetic lines profiles. 
For the coming years, the development of such models is strongly needed.


\medskip

We present here simultaneous observations of profile variations of sodium and \ha lines in \rrly itself. 
In Sect.\,2 we describe observation data and their reduction processes. 
Observations analyses including the sodium line velocity measurement are presented in Sect.\,3.
Atmospheric implications of sodium observations are discussed in Sect.\,4 and the dynamical structure of the atmosphere is described in Sect.\,5. 
Section\,6 is dedicated to shock wave dynamics.
Finally, we draw our conclusions in Sect.\,7.

\section{Observations and data analysis}

\subsection{Data acquisition}

In order to describe the atmosphere and shock wave dynamics, data were assembled from different sources. 
On the one hand, to achieve a complete movie of atmospheric layers over the whole pulsation cycle, we used time series spectra 
focused on Na\,D lines. 
On the other hand, for the shock wave study, since a 300\,s time resolution was required to complete velocity measurements, we reexamined previous published spectra.
Data used in this paper were obtained with three different spectrographs:

{\setlength{\tabcolsep}{3.5pt} 
\begin{table*}
\centering
\caption{Characteristics of the RR\,Lyr spectra. 
Columns provide the date of the night, the corresponding Julian Date, the telescope (Tel.), the observatory (Obs.), the attached spectrograph (Spectro.), 
its resolving power, its resolution element (RE) at $\lambda5896.92$, the spectral range, the mean S/N per pixel in the $\lambda$6630 region, 
the start and end of the pulsation phase $\varphi$ and the Blazhko phase $\psi$ for each night, the number $N$ of observed spectra, and finally the 
exposure time T$_{\textrm{exp}}$ used for the night.}
\label{longobslog}
\begin{tabular}{cccccccccccccc}
\hline
\hline
Night & JD  & Tel. & Obs. & Spectro.   & Resolving & RE & $\lambda_{\rm start} - \lambda_{\rm end}$ & S/N & $\varphi_{\rm start}$  & $\varphi_{\rm end}$ & $\psi$ & N & T$_{\textrm{exp}}$ \\
{\scriptsize (yyyy-mm-dd)} & {\scriptsize (-2\,400\,000)} &  &  & & power  & (pixel) &(\AA) &   &  & & & & (s)  \\ 
\hline
1994-08-03                  & 49568       & 193 & OHP & \textsc{Elodie} & 44\,000  &2.1 &$3\,800-6\,800$ & 78 & 0.08 & 0.09 & 0.39 & 1 & 420  \\
1996-08-02                  & 50298       & 193 & OHP & \textsc{Elodie} & 44\,000  &2.1 &$3\,800-6\,800$ & 60 & 0.08 & 0.09 & 0.08 & 1 & 480  \\
1997-06-03                  & 50603       & 193 & OHP & \textsc{Elodie} & 44\,000  &2.1 &$3\,800-6\,800$ & 84 & 0.08 & 0.09 & 0.89 & 1 & 300  \\
1997-08-05                  & 50666       & 193 & OHP & \textsc{Elodie} & 44\,000  &2.1 &$3\,800-6\,800$ & 32 & 0.08 & 0.09 & 0.50 & 1 & 300  \\
1997-08-09                  & 50670       & 193 & OHP & \textsc{Elodie} & 44\,000  &2.1 &$3\,800-6\,800$ & 17 & 0.08 & 0.09 & 0.60 & 1 & 480  \\
1997-08-31                  & 50692       & 193 & OHP & \textsc{Elodie} & 44\,000  &2.1 &$3\,800-6\,800$ & 69 & 0.08 & 0.09 & 0.17 & 1 & 300  \\
\hline
2013-10-12                  & 56578            & 152 & OHP & \textsc{aurelie} & 22\,700  &2.8 &$6\,500-6\,700$ & 86 & 0.04 & 0.04 & 0.91 & 1 & 300  \\
2013-10-16                  & 56582            & 35 & Chelles &\textsc{eShel I} &10\,500 &3.2 &$4\,300-7\,100$ &  75 & 0.08 & 0.09 & 0.03 & 1 & 300  \\
2014-05-01                  & 56778            & 35 & Foncaude&\textsc{eShel II}&10\,500 &4.3 &$4\,300-7\,100$ &  56 & 0.90 & 0.94 & 0.36 & 1 & 300  \\
2014-08-13                  & 56883            & 35 & Foncaude&\textsc{eShel II}&10\,500 &4.3 &$4\,300-7\,100$ &  43 & 0.90 & 0.94 & 0.04 & 7 & 300  \\
2016-07-18                  & 57588            & 35 & Chelles &\textsc{eShel I} &10\,500 &3.2 &$4\,300-7\,100$ &  69 & 0.08 & 0.09 & 0.00 & 1 & 300  \\
\hline
2017-04-07                  & 57851            & 35 & Chelles &\textsc{eShel I} &10\,500 &3.2 &$4\,300-7\,100$ &  76 & 0.04 & 0.42 & 0.75 & 21 & 900  \\
2017-04-08                  & 57852            & 35 & Chelles &\textsc{eShel I} &10\,500 &3.2 &$4\,300-7\,100$ &  78 & 0.80 & 0.18 & 0.77 & 18 & 900  \\
2017-04-09                  & 57853            & 35 & Chelles &\textsc{eShel I} &10\,500 &3.2 &$4\,300-7\,100$ &  74 & 0.56 & 0.85 & 0.80 & 16 & 900  \\
2017-04-10                  & 57854            & 35 & Chelles &\textsc{eShel I} &10\,500 &3.2 &$4\,300-7\,100$ &  60 & 0.37 & 0.50 & 0.83 & 7 & 900  \\
2017-04-11                  & 57855            & 35 & Chelles &\textsc{eShel I} &10\,500 &3.2 &$4\,300-7\,100$ &  80 & 0.20 & 0.47 & 0.85 & 5 & 900  \\
2017-04-12                  & 57856            & 35 & Chelles &\textsc{eShel I} &10\,500 &3.2 &$4\,300-7\,100$ &  86 & 0.92 & 0.22 & 0.88 & 6 & 900  \\
2017-04-17                  & 57861            & 35 & Chelles &\textsc{eShel I} &10\,500 &3.2 &$4\,300-7\,100$ &  77 & 0.65 & 0.02 & 0.00 & 6 & 900  \\
2017-05-17                  & 57891            & 35 & Chelles &\textsc{eShel I} &10\,500 &3.2 &$4\,300-7\,100$ &  72 & 0.08 & 0.09 & 0.75 & 3 & 600  \\
\hline
\end{tabular}
\end{table*}
}

\begin{itemize}

\item \textsc{eShel I} spectrograph:
We used an echelle spectrograph called \textsc{eShel} attached to an automated 35\,cm telescope at the Observatoire de Chelles \citetext{Lemoult et al. in prep.}.
The fiber-fed \textsc{eShel} spectrograph is described in \citet{2011IAUS..272..282T} and is built by Shelyak Instruments.\footnote{\url{http://www.shelyak.com}}
The spectral domain is spread over $4\,300-7\,100$\,\AA\ for a resolving power of $R=10,500$ at $\lambda5896.92$. 
The detector used is an Atik 460EX camera (Sony ICX694 CCD) on which the resolution element represents about 3.2 pixels.
The exposure time was 900\,s providing a mean S/N of 75.
A total of 47\,h of exposure time were gathered between April 7 and 18, 2017.
In order to cover a full cycle, we selected 79 spectra for a total of 20\,h of exposures.
These 79 spectra achieve a complete and continuous \rrlyr cycle with a maximum of 900\,s, \ie, a 2\,\% cycle step.
Observations were started at the Blazhko phase $\psi=0.77$ to finish at maximum Blazhko $\psi=0.00$ in a single Blazhko cycle. 

\smallskip
\item \textsc{eShel II} spectrograph: 
In August 2014, we observed with a similar echelle spectrograph \textsc{eShel} attached to a 35\,cm telescope at the Observatoire de Foncaude, France.
The detector was a CCD sensor Kodak KAF-3200ME. The resolution element represents about 4.3 pixels.
The aim was to collect mid-resolution spectra of RR Lyr over 45\,min around the outward atmosphere acceleration at $\varphi=0.91$. 
Eight 300\,s spectra were obtained over the spectral visible domain ($4300-7100$~\AA) and with a resolution power of about $10,500$ and a S/N of 50.

\smallskip
\item \textsc{Aurelie} spectrograph: 
On October 12, 2013, we used the \textsc{Aurelie} \citep{gillet94} spectrograph ($1\,200$~t/mm, Blaze $5\,000$~\AA\ with OG590 order filter) 
recorded with a EEV 42-20 CCD on T152 at the  Observatoire de Haute-Provence (OHP), France. In order to collect high-resolution spectra of RR Lyr,
600\,s spectra were acquired over the visible domain (6513-6713~\AA) and with a resolution power of about $22,700$ and a S/N of 86. The resolution element represents about 2.8 pixels.

\smallskip
\item \textsc{Elodie} spectrograph: Attached to the 193\,cm telescope at the OHP, it is described in \citet{1996A&AS..119..373B}.
Observations were performed in $1994-1997$ in the context of a survey led by D. Gillet and published in \cite{1999A&A...352..201C}.
We used some spectra from these observations to verify the Na\,D red component at different pulsation phases.
Typical exposure times were between $5-8$\,min, leading to a
signal-to-noise ratio (S/N) of about 57 per pixel for a resolving power of $R = 44,000$ on the whole visible domain ($3\,900-6\,800$~\AA).
\end{itemize}

The observations are summarized in Table \ref{longobslog}.

\subsection{Data reduction}

\textsc{Aurelie} data were reduced with the SpcAudace package \citep{spcaudace2017}, part of Audela\footnote{\url{http://www.audela.org}} software \citep{audela2012}. 
\textsc{eShel} spectra processing was done using a dedicated echelle sub-package of the Audela software too. \textsc{Elodie} spectrograph uses its own MIDAS pipeline \citep{1996A&AS..119..373B}.

These packages perform classical operations through pipelines, such as preprocessing (bias and dark subtraction, flat-fielding), masking of bad pixels, 
spectrum extraction, and wavelength calibration. Line profiles were corrected from instrumental response and heliocentric velocity. 
All spectra were normalized to the local continuum around the \ha line.

Wavelengths are presented in the stellar rest frame of \rrly using its actual radial velocity $V_\ast=-73.5$\,\kms with respect to the 
solar system barycenter \citep{chadid96}. 
$V_\ast$ is also called the $\gamma$-velocity. 
All wavelengths ($\lambda$) are given in angstr\"om (\AA).

\subsection{Ephemeris computation}

In this paper, the pulsation phase is denoted $\varphi$, where $\varphi=0.0$ corresponds to the luminosity maximum. 
The Blazhko phase is denoted $\psi$ and has its maximum at $\psi=0.0$ when the highest luminosity amplitude is observed during a Blazhko cycle.

The period and reference date of pulsation and Blazhko ephemeris evolve with time. 
Up-to-date parameters are thus required.
\cite{leborgne14} showed that the pulsation period varies alternately between two states, defined as a pulsation over  long (longer than 0.56684\,d) and  short (shorter than 0.56682\,d) periods. These states alternate with intervals of 13--16\,y. After a sudden change in period value, this interval remains constant for a few years. The physical origin of these two states is not known yet.

{\setlength{\tabcolsep}{4.4pt} 
\begin{table}[!h]
\caption{Ephemerides used for the pulsation and Blazhko phase computation.}
\centering
\label{ephemeris}
\begin{tabular}{c|ll|ll}
\hline
\hline
Epoch & \multicolumn{2}{c}{Pulsation} & \multicolumn{2}{|c}{Blazhko} \\
\hline
      & HJD0$_{P}$              & Period                      & HJD0$_{B}$   & Period \\
      & {\scriptsize (-2\,400\,000)}& $P_P$ (d) & {\scriptsize (-2\,400\,000)} & $P_B$ (d)  \\ 
\hline
1994  & 50\,456.7090\tablefootmark{a} & 0.5668427\tablefootmark{b} & 49\,631.312\tablefootmark{c} & 39.03\tablefootmark{b} \\
1996  & 50\,456.7090\tablefootmark{a} & 0.5668174\tablefootmark{b} & 49\,631.312\tablefootmark{c} & 39.06\tablefootmark{b} \\
1997  & 50\,456.7090\tablefootmark{a} & 0.5668174\tablefootmark{b} & 49\,631.312\tablefootmark{c} & 39.06\tablefootmark{b} \\
2013  & 56\,539.3428\tablefootmark{a} & 0.5667975\tablefootmark{b} & 56\,464.481\tablefootmark{d} & 39.0\tablefootmark{b} \\ 
2014  & 56\,914.5507\tablefootmark{a} & 0.5667975\tablefootmark{b} & 56\,881.627\tablefootmark{d} & 39.0\tablefootmark{b} \\
2017  & 57\,861.6319\tablefootmark{a} & 0.566793\tablefootmark{a}  & 57\,354.322\tablefootmark{d} & 39.0\tablefootmark{b} \\
\hline
\end{tabular}
\tablefoot{Maximum light HJD and periods are from:
\tablefoottext{a}{GEOS;\,}
\tablefoottext{b}{\citet{leborgne14};\,}
\tablefootmark{(c)}{\citet{1997A&A...319..154C};\,}
\tablefootmark{(d)}{This paper}.
}
\end{table}
}

\begin{figure*}
  \centering
  \includegraphics[width=\textwidth]{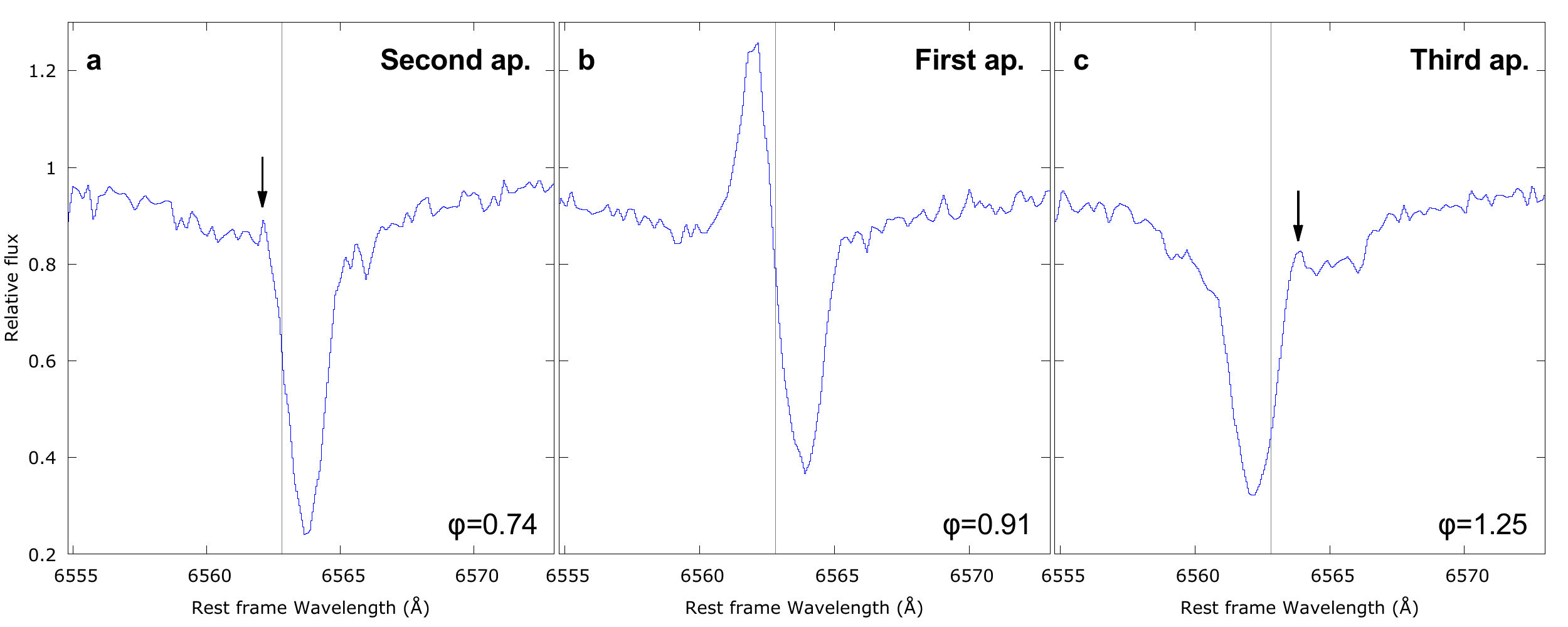}
  \caption{Comparison of the three apparitions of \ha line profile of \rrlyn. 
  \textbf{Panel a:} Second apparition at $\varphi=0.74$ represented with a spectrum dated on 2017-Apr-10.
  \textbf{Panel b:} First apparition at $\varphi=0.91$ represented with a spectrum dated on 2017-Apr-09.
  \textbf{Panel c:} Third apparition at $\varphi=1.25$ represented with a spectrum dated on 2017-Apr-08.
  Vertical line indicates the \ha line laboratory wavelength.}
  \label{plot_3_emissions}
\end{figure*}

We thus used the ephemerides provided by \citet{leborgne14} for the \textsc{aurelie} and \textsc{eShel II} observations made in 2013 and 2014.
For the 2017 data, a late epoch for the ephemeris provided by \citet{leborgne14}, the pulsation period $P_P$ was computed using the difference between the 
two closest $O-C$ minima from the GEOS RR\,Lyr web database\footnote{\url{http://rr-lyr.irap.omp.eu}} and close to our 2017 observations. 
The reference HJD maximum light (HJD0$_P$) was also provided by this database.
A summary of the  ephemerides used for the different data sets is provided in Table\,\ref{ephemeris}.

\medskip

The determination of the Blazhko phase is more problematic since the Blazhko period $P_B$ does not follow the variation in the pulsation period, 
but presents rather erratic changes \citep{leborgne14}.
Moreover, during 2014, the photometric $O-C$ were historically low (close to nil), preventing an easy Blazhko period computation.

Therefore, in order to find maximum light amplitude dates HJD0$_B$, we developed an innovative method, which will be described in a forthcoming paper (Gillet et al.
in prep.), based on the equivalent width (EW) and main 
shock velocities driven by our many years of experience in performing spectral observations. 
For each photometric maximum ($\varphi=0$) corresponds to a spectroscopic maximum ($\varphi=0.91$) where \ha presents an emission line. 
Consequently, |EW| and $|V_{\textrm{shock}}|$ (defined in Eq.~\ref{vshock1}) reach high values. 
They  reach a maximum when the light amplitude $O-C$ correspond to their extrema, thus pointing out \bla maxima.
We expect an uncertainty of $\pm 2$\,days on HJD0$_B$ and of $\pm 0.2$\,d on $P_B$. 
Ephemerides were computed involving the period $P_{B}=39.0$\,d \citep{leborgne14}.
A summary of the Blazhko ephemeris used for the different data set is provided in Table\,\ref{ephemeris}.

\section{Observations of \ha and sodium in \rrlyn}

We analyzed profile variations for three spectral lines: \ha, He, and Na\,D1 during a complete pulsation period. 
This last line is a resonance line that, following \cite{fokin97}, is formed in the deep layers of the photosphere, 
while the core of the \ha and helium lines are formed well above the photospheric layers,  hereafter called the upper atmosphere. 
Thus, these three lines, by their different formation heights, are able to probe and to follow the propagation of the waves 
in the atmosphere along the pulsation cycle.

Because one observing night is not long enough to obtain spectra during the whole pulsation period ($P_P\simeq13$\,h\,36\,min), we had to recompose the data of 11 consecutive nights. It was shown by \cite{gillet88} that lines could vary depending on the Blazhko phase with a period of $P_B\simeq39$\,d. Thus, as 55 spectra out of 79 were obtained over three days, on this timescale the Blazhko effect must be secondary in the framwork of our data. Consequently, the use of data obtained in different periods will not change appreciably the different phases of the atmospheric dynamics and the effects of the shock wave propagation on these different phenomena, including their impact on the turbulence amplification.

\subsection{\ha line evolution}

\begin{figure}
  \centering
  \includegraphics[width=0.97\hsize]{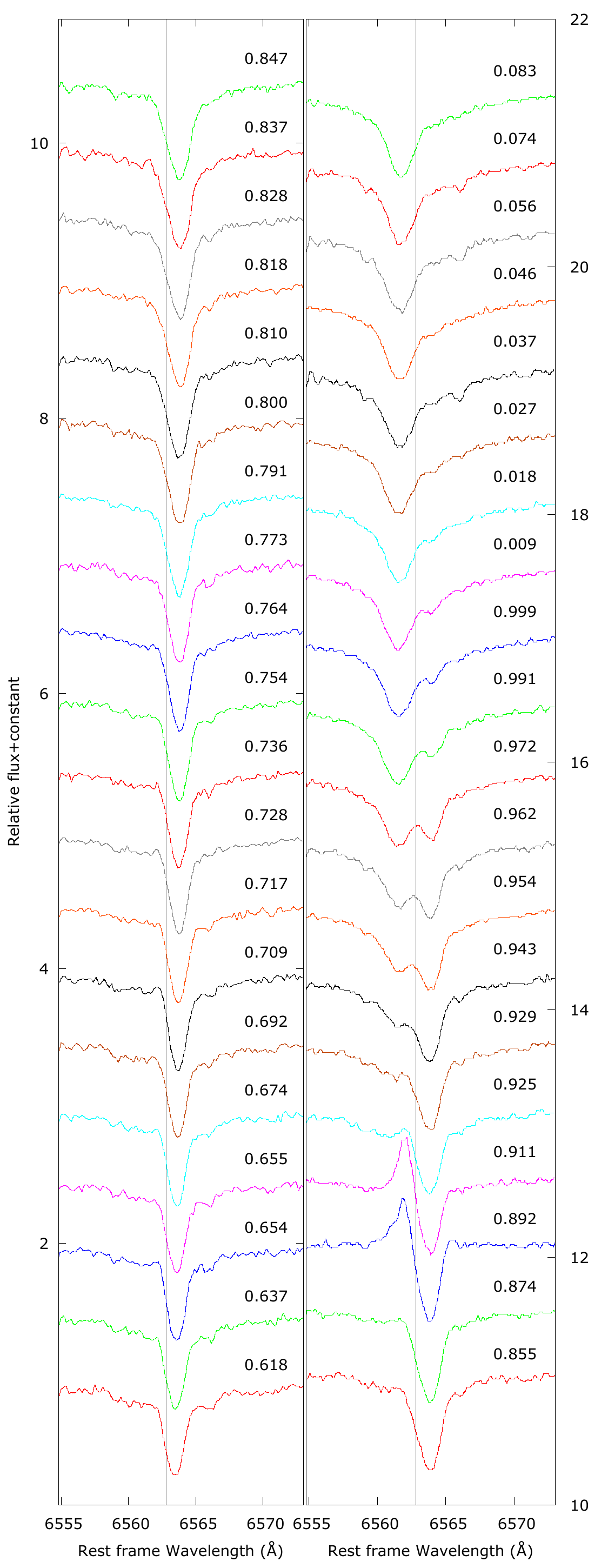}
  \caption{
    Evolution of the \ha line profile based on \rrly observations obtained in April 2017.
    The pulsation phase varies from $\varphi=0.618$ (bottom left) to $\varphi=0.083$ (top right), while the Blazhko phase $\psi$ overlays the interval 0.77-0.03. 
    The exposure time is 900\,s, but the time step is shorter.
    Weak absorption features are telluric lines.
    The vertical line indicates the \ha line laboratory wavelength.}
  \label{ha_film_201704-2}
\end{figure}

\begin{figure}
  \centering
  \includegraphics[width=0.97\hsize]{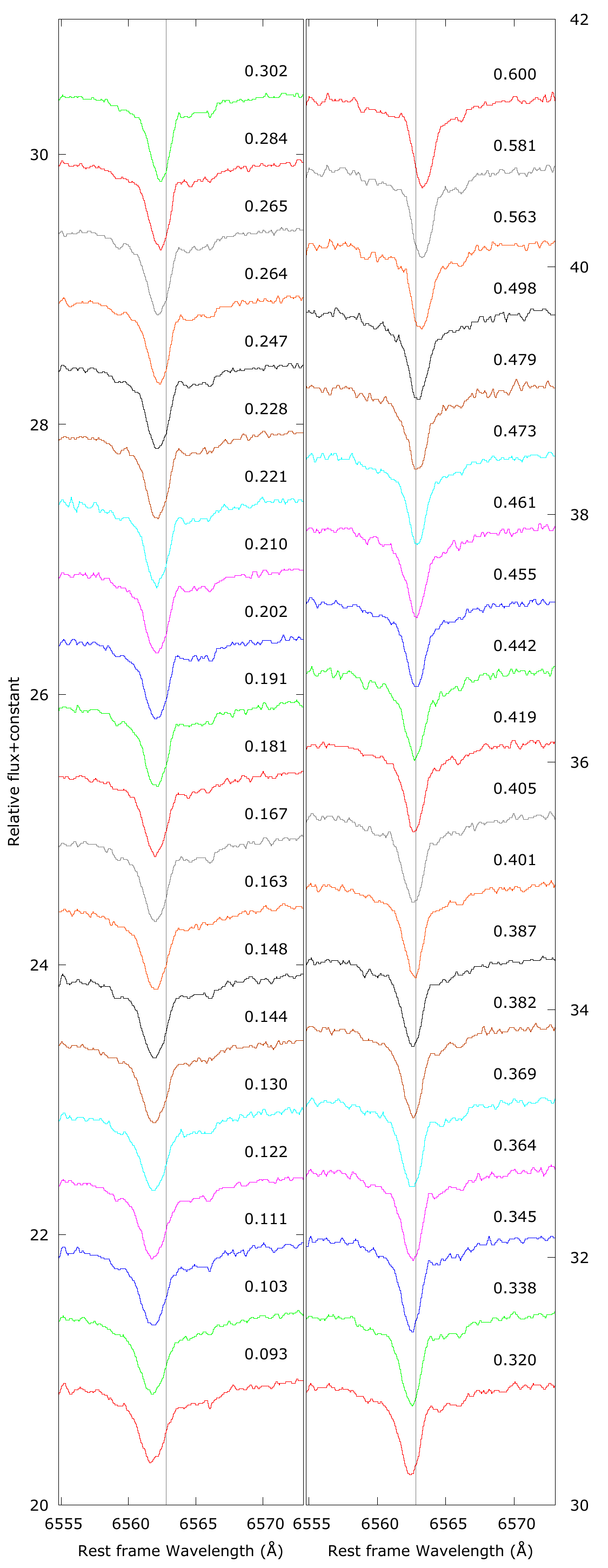}
  \caption{
    Evolution of the \ha line profile based on \rrly observations obtained in April 2017.
    The pulsation phase varies from $\varphi=0.093$ (bottom left) to $\varphi=0.600$ (top right), while the Blazhko phase $\psi$ overlays the interval 0.77-0.03. 
    The exposure time is 900\,s, but the time step is shorter.
    Weak absorption features are telluric lines.
    The vertical line indicates the \ha line laboratory wavelength.}
  \label{ha_film_201704-1}
\end{figure}

The three successive appearances of hydrogen emission \citep{preston11} have been  observed in detail.
Between $\varphi=0.892-0.929$, just before the luminosity maximum, an intense blueshifted emission is present, the well-known ``first apparition.'' 
A smaller red bump is detectable from $\varphi=0.637$ to $\varphi=0.736$. 
Panel a of Fig.\,\ref{plot_3_emissions} zooms in on this ``second apparition,'' which is the weak hump at the top of the blue line wing around $\varphi\approx0.73$.
Finally, another small red bump corresponding to the ``third apparition'' is detected for $\varphi=0.181-0.387$.
These three apparitions are compared in Fig.\,\ref{plot_3_emissions}. 
While only one line profile is plotted for each apparition in this figure, they are observed in several consecutive spectra (see Figs.\,\ref{ha_film_201704-2} 
and \ref{ha_film_201704-1}). 

The blueshifted emission intensity corresponding to the first apparition (Fig.~\ref{plot_3_emissions}, panel b) represents up to 30\% of the continuum, depending 
on the Blazhko phase. 
The duration of the emission of the \ha line for the first apparition is between 15\,min and 40\,min, \ie, less than 5\% of the pulsation cycle. 
This phenomenon is rather short compared to the second and third apparitions (Fig.\,\ref{plot_3_emissions} panels a and c, respectively). 
Their durations are respectively about 1.5\,h and 3\,h, \ie, 11\% and 20\% of the pulsation period. 

The second apparition was first observed in \rrly by \cite{gillet88}. 
They measured a 30\,min duration in nonconsecutive observations. 
This emission may depend on the pulsation cycle and requires observations over several cycles to be clearly defined. 

On the other hand, the third apparition was first reported in \rrly itself by \cite{gillet2017a}, where the emission intensity and duration are consistent with our observations. 
However, these two apparitions intensities represent only a few percents of the continuum.

Figures\,\ref{ha_film_201704-2} and \ref{ha_film_201704-1} show the evolution of the \ha line profile over the full pulsation cycle. 
They start at the   bottom left and go toward the top right. 
The \ha line becomes redshifted at $\varphi=0.618$. 
It is followed by a line doubling starting at $\varphi=0.943$ and ending at $\varphi=1.027$, \ie, $\sim8\%$ of the pulsation cycle. 
Then the \ha line remains blueshifted until $\varphi=1.461$, where it reaches its laboratory wavelength, and then becomes  redshifted again.
The shock intensity depends on the \bla phase: the more  important it is, the more visible the \he emission lines become.

\subsection{Helium line evolution}

\begin{figure}[!ht]
  \centering
  \includegraphics[width=0.97\hsize]{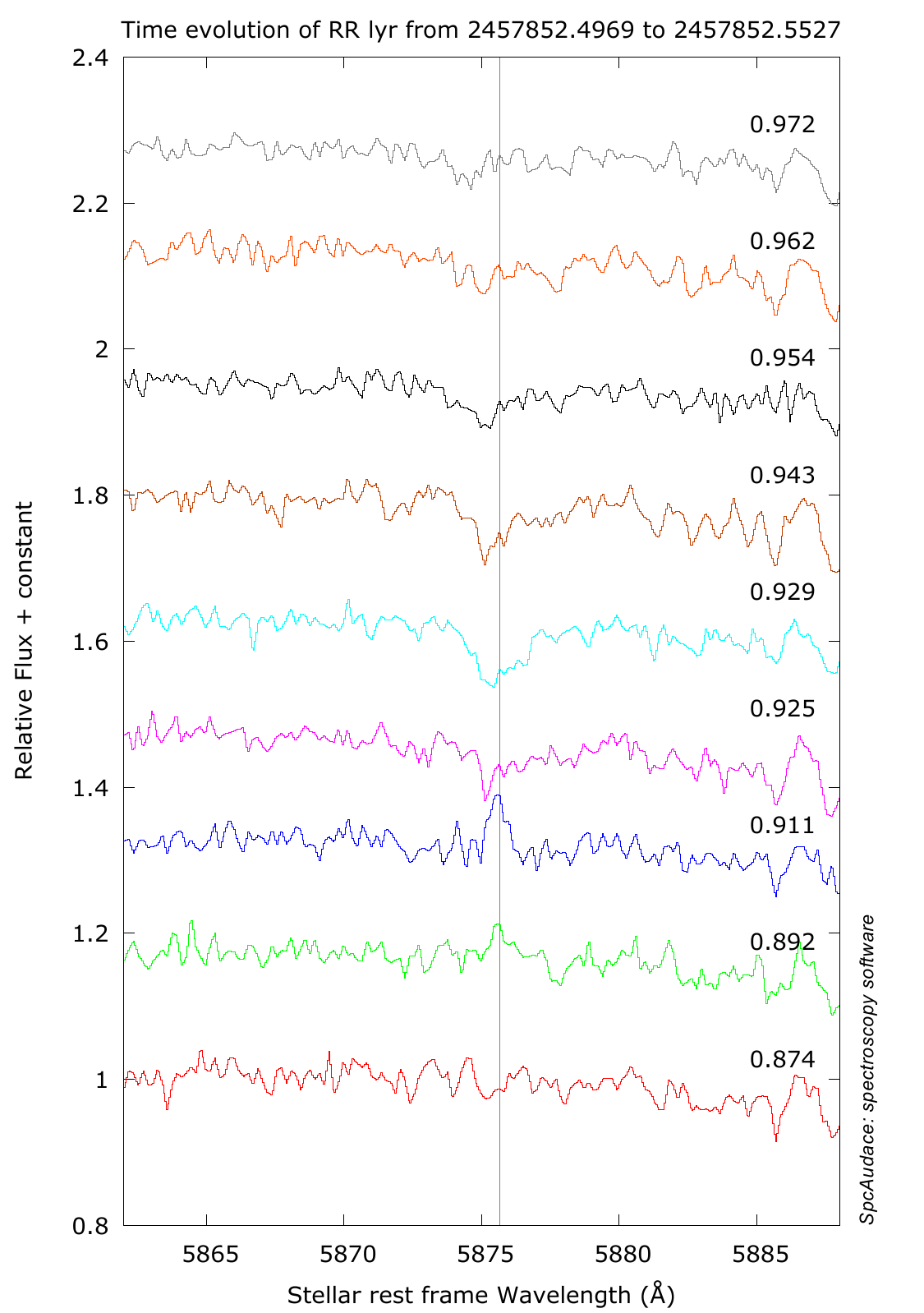}
  \caption{Apparition of the \he $\lambda 5876$ emission line at $\varphi=0.911,$ \ie, when the \ha line is also in emission. 
The vertical line indicates the \he line laboratory wavelength.}
  \label{he5876emission}
\end{figure}

Figure\,\ref{he5876emission} shows the \he $\lambda 5876$ line in emission at $\varphi=0.911$, \ie, the same phase of the first apparition of the \ha line. 
This emission was first reported in \rrly by \cite{GFL2013}.
Its intensity is 8\% of the continuum and the line velocity reaches $-6.5\pm0.8$\kms, which confirm the \cite{GFL2013} measurements.
While this emission is a short phenomenon of  less than 30\,min, \ie, $\sim3$\% of the pulsation period, it is followed by an absorption shape 
between $\varphi=0.925-0.954$. 
We note  that the low value $\textrm{S/N}_{\textrm{He line}}\cong3$ prevents us from performing a more detailed analysis of such a weak feature.
No emission of the \he $\lambda 6678$ line is observed at the same phase. 
This is consistent with the Blazhko phase $\psi=0.77$ for this observation on the night  of April 8--9, 2017. 
Indeed, this Blazhko phase corresponds to a strong shock, but not an exceptional one. 
Stronger shocks are expected around $\psi=0$ \citep{gillet2013}.

\subsection{Sodium line evolution}

\begin{figure}
  \centering
  \includegraphics[width=0.97\hsize]{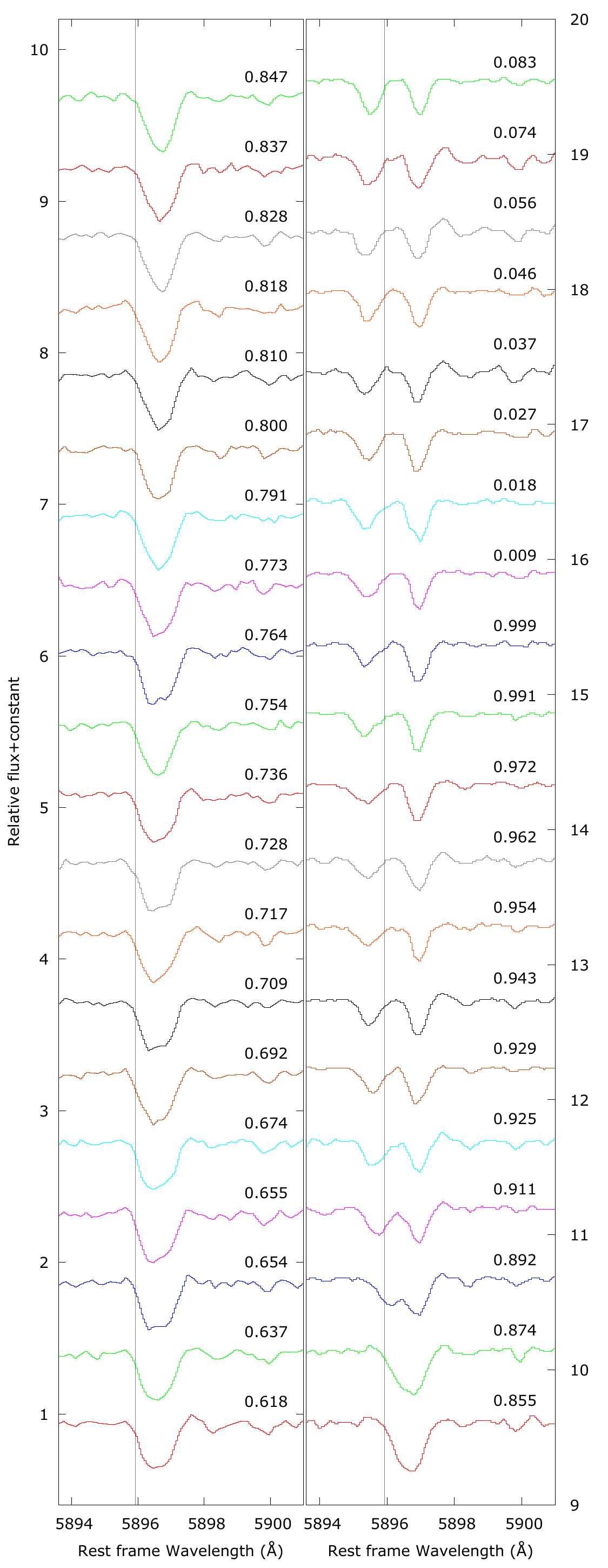}
  \caption{
    Evolution of the D1 sodium line profile ($\lambda$5895.92) based on \rrly observations obtained in April 2017.
   The pulsation phase varies from $\varphi=0.618$ (bottom left)  to $\varphi=0.083$ (top right), while the Blazhko phase $\psi$ overlays the interval 0.77-0.03. 
   The exposure time is 900\,s, but the time step is shorter.
   There is a line doubling that  stays over 75\% of the cycle.
   The vertical line indicates the sodium line laboratory wavelength.}
  \label{na_film_201704-1}
\end{figure}

\begin{figure}
  \centering
  \includegraphics[width=0.97\hsize]{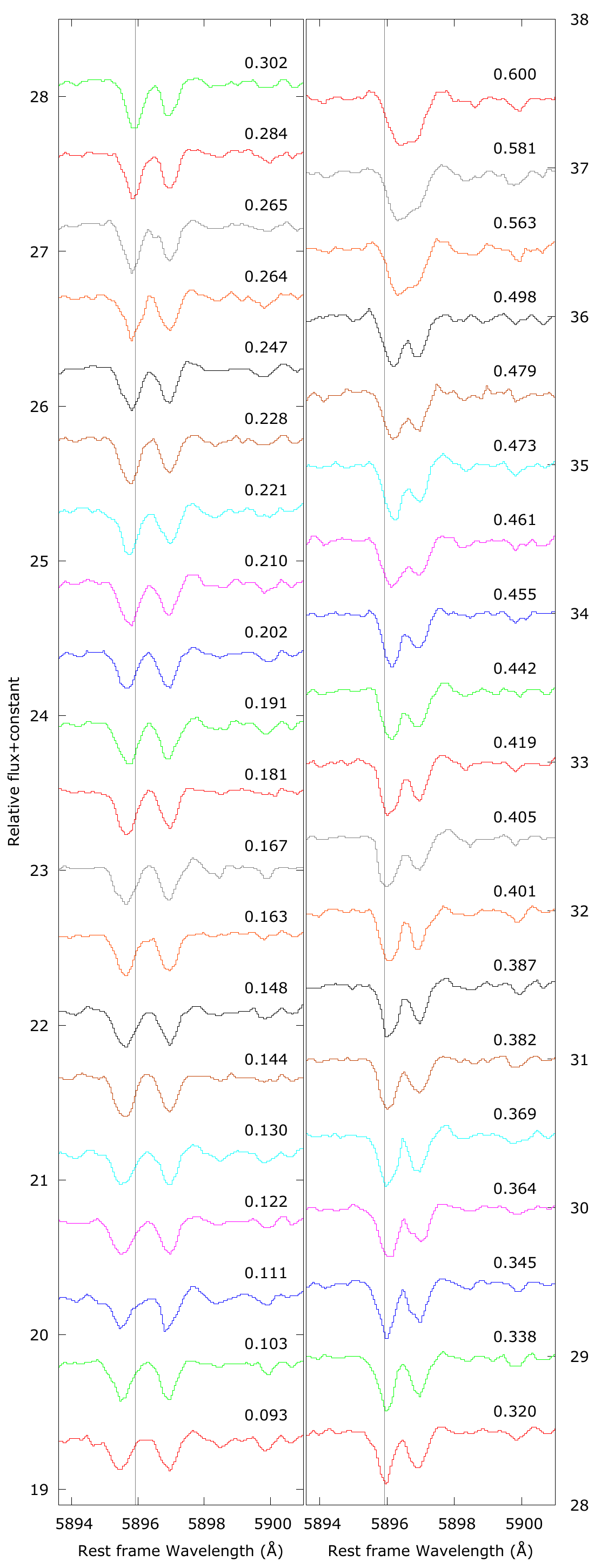}
  \caption{
    Evolution of the D1 sodium line profile ($\lambda$5895.92) based on \rrly observations obtained in April 2017.
   The pulsation phase varies from $\varphi=0.093$ (bottom left) to $\varphi=0.600$ (top right),  while the  Blazhko phase $\psi$ overlays the interval 0.77-0.03. 
   The exposure time is 900\,s, but the time step is shorter.
   There is a line doubling which remains over 75\% of the cycle.
   The vertical line indicates the sodium line laboratory wavelength.}
  \label{na_film_201704-2}
\end{figure}

Figures\,\ref{na_film_201704-1} and \ref{na_film_201704-2} show the evolution of the D1 sodium line profile ($\lambda$5895.92) over the full pulsation cycle.
Two strong telluric lines are blanketing the center of the D2 sodium component at $\lambda 5889.95$ line \citep{hobbs78}. 
Consequently, we focused our analysis on the D1 sodium component at $\lambda$5895.92 line. 
This line is denoted  Na\,D in the paper.
Figures\,\ref{na_film_201704-1} and \ref{na_film_201704-2} show the evolution of the D1 line profile over the full pulsation cycle.
 
A line doubling starts somewhere for $\varphi\in[0.892;0.925]$, \ie, very close to the pulsation phase of the first apparition of the \ha line. 
It remains visible along $\varphi\in[0.855;1.600]$, \ie, over 75\% of the pulsation cycle. 
The origin and evolution of these two components of the D1 line profile are analyzed in detail in Sects.\,4 and 5.


\subsection{Line velocity evolution}

\begin{figure}[!ht]
  \centering
  \includegraphics[width=\hsize]{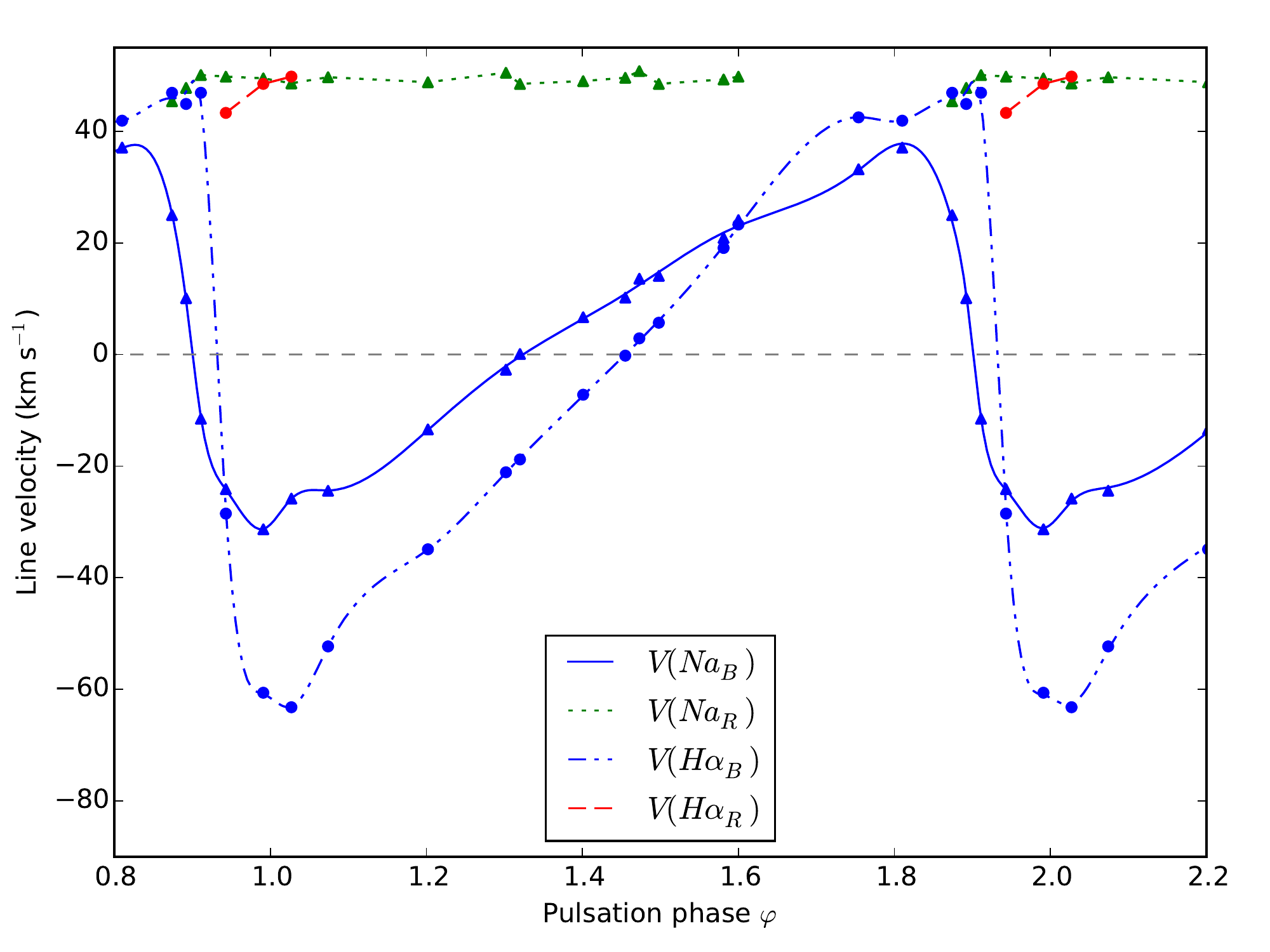}
  \caption{Curves of velocity with respect to the stellar rest frame of the sodium and \ha over the pulsation phase $\varphi=0.8$ to $\varphi=2.2$. 
   Shown are the blue components,  $V(Na_B)$ (blue filled line) and $V(H\alpha_B)$ (blue dot-dashed line), and  their respective red components,
     $V(Na_R)$ (green dotted line) and $V(H\alpha_R)$ (red dashed line). The mean uncertainty is $\pm 0.6$ \kmsn.}
  \label{na_ha_velocities}
\end{figure}

From the whole time series composed of 79 spectra, we selected 19 spectra that led us to describe every step of the layer motion evolution 
during one pulsation cycle.
In order to probe atmospheric dynamics we extracted the radial velocities of  the sodium and the \ha absorption lines. 
Measurements were done with respect to the stellar rest frame.
Line center measurements were obtained by Gaussian fitting or first moment of the line profile,  depending on the line shape. 
The mean uncertainty is $\pm0.6$\kmsn. 
These Doppler velocities are listed in Table\,\ref{vabs2017} and presented in Fig.\,\ref{na_ha_velocities} where curves are computed with a cubic spline.

The sodium blueshifted component has a very perturbed behavior, alternating receding and ascending movements. 
However, the redshifted velocity component remains  constant around 50\kmsn. 
A detailed analysis is presented in Sect.\,4.

{\setlength{\tabcolsep}{2.5pt} 
\begin{table}[!ht]
\caption{Radial velocity measurements with respect to the stellar rest frame of sodium and H$\alpha$ absorption lines within the pulsation phase. 
The mean uncertainty computed from line center measurements is $\pm0.6$\kmsn.
}
\label{vabs2017}
\centering
\begin{footnotesize}
\begin{tabular}{>{$}c<{$} | >{$}c<{$} >{$}c<{$} | >{$}c<{$}}
\hline\hline
\textrm{Phase} & \multicolumn{3}{c}{\textrm{Line velocity (km\,s}$^{-1}$)} \\ \hline
\varphi & \textrm{Na (blue)} & \textrm{Na (red)} & \textrm{H}\alpha\textrm{ (abs)} \\ \hline
0.874 & +25.0 & +45.4 & +46.9 \\
0.892 & +10.1 & +47.8 & +44.9 \\
0.911 & -11.5 & +50.1 & +46.9 \\
0.943 & -24.1 & +49.8 & -58.5\tablefootmark{a}~~+43.3\tablefootmark{a} \\
0.991 & -31.3 & +49.5 & -60.6\tablefootmark{a}~~+48.5\tablefootmark{a} \\
1.027 & -25.8 & +48.6 & -63.2\tablefootmark{a}~~+49.8\tablefootmark{a} \\
1.074 & -24.3 & +49.7 & -52.3 \\
1.202 & -13.4 & +48.8 & -34.9 \\
1.302 & -2.7  & +50.5 & -21.1 \\
1.320 & +0.1  & +48.5 & -18.8 \\
1.401 & +6.7  & +49.0 & -7.2 \\
1.455 & +10.2 & +49.6 & -0.2 \\
1.473 & +13.6 & +50.8 & +2.9 \\
1.498 & +14.1 & +48.5 & +5.7 \\
1.581 & +20.9 & +49.3 & +19.1 \\
1.600 & +24.1 & +49.8 & +23.3 \\
1.618 & \multicolumn{2}{c|}{+34.9\tablefootmark{b}} & +27.8 \\
1.754 & \multicolumn{2}{c|}{+33.2\tablefootmark{b}} & +42.5 \\
1.810 & \multicolumn{2}{c|}{+37.1\tablefootmark{b}} & +41.9 \\
\end{tabular}
\end{footnotesize}
\tablefoot{
\tablefoottext{a}{\ha line doubling;\,}
\tablefoottext{b}{Na single line.}
}
\end{table}
}

Like all spectral lines, the \ha absorption line evolution alternates ascending and receding movements over the whole pulsation cycle.
Near $\varphi=0.892$, the first emission arises. This is followed by a brief line doubling episode  ($\varphi=0.943-1.027$), during which the extremal blue ($B$) and red ($R$) radial velocities of these two components are $V(H\alpha_B)=-63.2$\kms 
and $V(H\alpha_R)=+49.8$\kmsn.
At $\varphi=1.037$, the \ha line starts a progressive redshift from its maximum blueshifted velocity value $-54.6$\kms to finish at $+46.9$\kms near $\varphi=1.874$.
The \ha line dynamics is studied in detail in Sect.\,4.


\section{The very redshifted sodium component}

Our observations (Figs.\,\ref{na_film_201704-1} and \ref{na_film_201704-2}) show that a very redshifted component is present within the Na\,D line 
profiles near $\lambda$5897. 
It is clearly visible from $\varphi=0.911$ to $\varphi=0.302$, \ie, when it is not blended with the variable pulsational sodium component. 
Between 1994 and 2017, its velocity remains constant near $+50.3\pm1.2$\kms (see Table\,\ref{na_measurements}). 
The shape of this component also seems  constant, \ie, it is not dependent on the pulsation phase. 
However, the line profiles of other metallic lines such as \ion{Fe}{ii} $\lambda4923.921$ and \ion{Mg}{i} $\lambda5183.604$ \citep{chadid96} as well as Balmer lines 
\citep{preston65} observed at  a resolving power of up to $R=44\,000$ do not present such a very redshifted component as observed for Na\,D lines. 
This points toward a nonphotospheric origin of this very redshifted Na\,D line component.

{\setlength{\tabcolsep}{4pt} 
\begin{table}[!ht]
\caption{Measurements of the very redshifted component within the Na\,D line profiles: $V_{\textrm{radial}}$ by Gaussian fitting  
with respect to the stellar rest frame, FWHM corrected from instrument function, and EW at $\varphi=0.000-0.083$ from the 1994 -- 2017 spectra. 
}
\label{na_measurements}
\centering
\begin{footnotesize}
\begin{tabular}{c >{$}c<{$} >{$}c<{$} >{$}c<{$} >{$}c<{$}}
\hline\hline
Date & \varphi & V_{\textrm{radial}} & \textrm{FWHM} & \textrm{EW} \\
(yyyy-mm-dd) &  & (\textrm{km\,s}^{-1}) & (\mbox{\AA}) & (\mbox{\AA}) \\ \hline
1994-08-03\tablefootmark{a} & 0.083 & 49.3 & 0.168 & 0.1554 \\
1996-08-02\tablefootmark{a} & 0.083 & 50.6 & 0.189 & 0.1541 \\
1997-06-03\tablefootmark{a} & 0.083 & 50.2 & 0.154 & 0.1580 \\
1997-08-05\tablefootmark{a} & 0.083 & 51.2 & 0.199 & 0.1565 \\
1997-08-09\tablefootmark{a} & 0.083 & 51.7 & 0.205 & 0.1558 \\
1997-08-31\tablefootmark{a} & 0.083 & 50.7 & 0.166 & 0.1652 \\
2013-10-16\tablefootmark{b} & 0.083 & 51.5 & 0.160 & 0.1443 \\
2014-05-01\tablefootmark{c} & 0.008 & 50.4 & 0.197 & 0.1686 \\
2016-07-18\tablefootmark{b} & 0.005 & 48.4 & 0.206 & 0.1652 \\
2017-05-17\tablefootmark{b} & 0.034 & 48.5 & 0.171 & 0.1543 \\
\hline
Mean \& RMS & - & 50.3\pm1.2 & 0.182\pm0.018 & 0.1577\pm 0.0070 \\
\hline
\end{tabular}
\tablefoot{Spectrograph used for observation:
\tablefoottext{a}{\textsc{Elodie};\,}
\tablefoottext{b}{\textsc{eShel}~I;\,}
\tablefootmark{(c)}{\textsc{eShel}~II}.
}
\end{footnotesize}
\end{table}
}

The maser emissions SiO or OH  are considered as proof of the presence of a circumstellar envelope. 
Nevertheless, until now this type of emission has never been detected for RR\,Lyr. 
In addition, microwave CO lines, which could be the fossil remnants of an ancient shock ejection, have not been detected. 
Moreover, RR\,Lyr is not known to have strong mass loss due to spectacular shock ejections like those occurring in unstable red giant 
envelopes \citep{tuchman78, tuchman79}. 
As a result, it is unlikely that RR\,Lyr has a circumstellar envelope.
The most likely explanation of this redshifted component is an interstellar medium (ISM) origin. 
It would be due to the absorption by interstellar dust and gas on the line of sight of \rrlyn.

The sodium splitting is seen in only 75\% of the pulsation cycle, but not in the other 25\%.
The stellar component merges with the static ISM component during $\varphi$ interval $0.563-0.874$ since the Na stellar component is redshifted at $\varphi=0.338-0.892$.

The corresponding large observed heliocentric velocity ($V_{\rm helio}=+50.3-73.5=-23.2$\kms, where $V_\ast=-73.5$\kmsn) of the redshifted component 
is consistent with an interstellar origin because, as RR\,Lyr has a low galactic latitude ($b=12$\,\ensuremath{^\circ}), the radial velocity of the 
expected interstellar material could be large due to small projection effect.
The observed FWHM of the red component remains approximately constant between 1994 and 2017 around $0.182\pm0.02$\,\AA, \ie, $9.3\pm0.9$\kms (see Table\,\ref{na_measurements}). 
This line width corresponds to the sum of a thermal part and a turbulent part occurring within the ISM gas. 
It is consistent with the FWHM observed on the high-resolution spectral profiles of the interstellar \ion{Na}{I}\,D line toward 80 early-type stars located in the local interstellar medium \citep{welsh1994}.

According to the GAIA database, RR\,Lyr has a parallax of $3.64\pm0.23$\,mas, \ie, a distance of 895\ light-years (ly) \citep{gaia2016}.
The Sun is located in a low-density region of the interstellar medium (about 0.05 atoms per cubic centimeter of neutral hydrogen, a near-perfect vacuum) 
called the Local Bubble \citep{frisch2011}. 
This gigantic asymmetric cavity of 330 to 490 ly in diameter has a wall of denser matter at its surrounding. 
It is believed that the Local Bubble has been carved by fast stellar winds and supernova explosions within the last 10\,Ma. 
Consequently, this cavity is partially filled with hot gas. 
RR\,Lyr is located in another expanding cavity called Loop III \citep{kun2007}. 
Between them, at about 200\,ly from RR\,Lyr,  Loop III and the Local Bubble are separated by a wall of about 150\,ly in size, of colder and denser neutral gas. 
Because the gas density within the cavities is at least 10 times lower than the average ISM density of the rest of the Milky Way's spiral disk, we can 
expect that the very redshifted sodium component is formed within the Local Bubble wall. 

\medskip

\cite{gillet2017a} have performed a single observation (2017-03-26) of the sodium line at $\varphi=0.227$. 
The profile shows a line doubling. 
Because the two components are similar (same FWHM and depth) it was thought that they were formed in a same atmosphere. 
Consequently, \cite{gillet2017a} suggested that the sodium doubling at $\varphi=0.227$ is due to the presence of a shock within the RR\,Lyr atmosphere. 
In the present paper we present a series of 79 spectra sampled approximately every 15\,min over the whole pulsation cycle, so we are able to show that the 
radial velocity of this component, when it is clearly visible, remains stable during the whole pulsation cycle for all line profile parameters. 
Consequently, we suggest that the redshifted component of the sodium profile has an interstellar origin. 
As a result, there would be no large velocity ($+50.3\pm1.2$\kmsn) sodium layer falling on the atmosphere around the pulsation phase $\varphi\sim0.30$, 
which would enhance the atmospheric compression causing the appearance of the third \ha emission.

\section{Dynamical structure of the atmosphere}

The pulsation of RR\,Lyr concerns essentially the photosphere and the atmospheric layers, which represent only a small percent of the mass of the star.    

Here we consider the main dynamic phenomena occurring during a pulsation period of RR\,Lyr of average amplitude. 
It is assumed to take place  about  halfway between the minimum and the maximum Blazhko, \ie, around the Blazhko phase $\psi=0.75$. 
The pulsation amplitude is minimum at Blazhko minimum and maximum at Blazhko maximum. 
The amplitude gradually increases between Blazhko minimum and Blazhko maximum. 
However, it is known that appreciable fluctuations are observed from one cycle to another. 
The Kepler-amplitude $K_p$ of the light curve of RR Lyr (see Figs.\,2 and 3 in \citealt{gillet2013}), which is correlated to the intensity of 
the main shock, strongly depends on the previous pulsation cycle. 
As a result, this random behavior  induces appreciable effects on the atmospheric dynamics, such as changes in observed line profiles. 
For example, helium lines can be in emission or not. 
Apart from these fluctuations, observations and models show that the dynamics of the pulsation follows a general reproducible scenario from one cycle to another. 
In this section we decompose these atmospheric phenomena into six main steps  that characterize the different dynamic processes occurring during a pulsation cycle (see Table~\ref{tocsection4}).

\begin{table}[!ht]
\caption{Decomposition of phenomenological steps of \rrly atmospheric dynamics during a typical pulsation cycle (see Sect.\,3).}
\label{tocsection4}
\centering
\begin{footnotesize}
\begin{tabular}{c>{$}c<{$}l}
\hline\hline
Step & \varphi\ \mathrm{interval} & \hfill Phenomenon \hfill \\ \hline
1 & 0.874-0.892 & Emergence of the main shock \\
2 & 0.892-0.929 & Radiative shock wave phase \\
3 & 0.320 & Maximum expansion of the sodium layer \\
4 & 0.455 & Maximum expansion of the \ha layer \\
5 & 0.320-0.455 & A two-step infalling motion \\
6 & \mathrm{around\ 0.36} & A strong phostospheric compression \\
\hline
\end{tabular}
\end{footnotesize}
\end{table}

\subsection{Phase 1 (0.874-0.892): emergence of the main shock}

Near the pulsation phase $\varphi=0.89$, a blueshifted \ha emission component suddenly appears within the \ha profile (Fig.\,\ref{ha_film_201704-2}). 
When the \ha emission reaches its maximum intensity ($\varphi=0.911$), \ion{He}{i} and \ion{H}{ii} lines can also be in emission if the light curve amplitude, 
hence the main shock, is large enough. 
Helium emission lines are not easily observed because their intensity is low (Fig.\,\ref{he5876emission}). 
Consequently, their exhaustive detection requires spectra with a high S/N, \ie, the use of large-diameter telescopes.

As noted in the introduction, RR\,Lyrae models that consider an atmosphere and shocks are rare. 
\cite{fokin97} presented in detail two models of \rrlyn\ that  give the same fine atmospheric structure (see Appendix\,A). 
At the end of the global compression of the star near the phase $\varphi=0.80$, the $\kappa-\gamma$ mechanism associated with the H-ionization zone
provokes a local overpressure, followed by the generation of a wave \itv{w1} that quickly breaks into the shock wave \itv{s1}. 
At approximately the same phase the second helium ionization zone, which is located far below the photosphere, also generates by a similar mechanism, 
a compression wave \itv{w2} which quickly turns into the shock wave \itv{s2}. 

As shown by \cite{fokin97}, these two waves propagate first toward the interior of subphotospheric layers. 
 In approximately the phase interval 0.75-0.90, the whole atmosphere falls on the star with an infall velocity exceeding that of the waves. 
Thus, all waves (or shocks) are first receding for the observer (Eulerian coordinates), but their motion is expanding outward in the Lagrangian frame. 
After  phase 0.90, the Mach number of shocks becomes as high as 15--25 and with the deceleration of the atmosphere during its downward motion, 
the shock starts to propagate outward in the Eulerian frame.

From the sodium and \ha line profiles reported in this paper (see Figs.\,\ref{ha_film_201704-1}, \ref{ha_film_201704-2}, \ref{na_film_201704-1}, and 
\ref{na_film_201704-2}), it is not possible to distinguish independently the shock waves \itv{s1} and \itv{s2}. 
Since the latter are formed well below the photosphere ($T\sim40,000$\,K), the optical depth is certainly too large for the \ha profile to be affected.

\subsection{Phase 2 (0.892-0.929): the shock is radiative}

For the observations presented in this paper, an \ha emission occurs at the pulsation phase $\varphi=0.892$. 
The emission is visible during a short time interval from $\varphi=0.892$ to $\varphi=0.929$, \ie, during 5\% of the pulsation cycle. 
The intensity of the shock is thus important enough to produce emission lines: it is a radiative shock.

The \ha emission is formed just behind the shock front within the radiative shock wake. 
This region is very narrow, typically a few kilometers depending on the shock velocity and physical condition within the atmosphere \citep{fadey04}. 
At what velocity (or Mach number) does the shock become radiative? 

At $\varphi=0.892$, the emission is blueshifted (and defined by $V_{e1}=-44.4$\kmsn, Fig.\,\ref{ha_film_201704-2}).
It is possible to obtain an estimate of the shock front velocity from the velocity of the blueshifted \ha emission. 
The gas layers emitting the Balmer line radiations are located at the rear of the shock front in the hydrogen recombination zone. 
We use shock models of \citet{fadey04}, which consider the structure of steady-state plane-parallel radiative shock waves propagating through 
a partially ionized hydrogen gas of pre-shock temperature $T_1 = 3\,000$\,K and pre-shock densities between $\rho_{1}=10^{-9}$\;g\,cm$^{\rm -3}$ 
and $10^{-12}$\;g\,cm$^{\rm -3}$. 
Thus, in the frame of the observer, the gas flow velocity in the hydrogen recombination zone is roughly one-half of the shock wave velocity:

\begin{equation}
\frac{V_{\textrm{shock}}}{V_{\textrm{recomb}}}\cong 2.00\pm0.11
\label{recomb}
\end{equation}

From these models it is also possible to estimate the ratio of $V_{e1}$ (denoted $\delta V$ in \citealt{fadey04}) of the hydrogen first emission component Doppler 
shift to the gas flow velocity $V_{\textrm{recomb}}$ in the hydrogen recombination zone:

\begin{equation}
\frac{V_{e1}}{V_{\textrm{recomb}}} \cong 0.70\pm0.03
\label{doppler}
\end{equation}

\noindent 
Combining these two equations, we obtain

\begin{equation}
V_{\textrm{shock}}(\varphi\simeq 0.90) \cong 3V_{e1} \cong 3c\frac{(\lambda_{e1}-\lambda_{0})}{\lambda_{0}} \,\pm0.16V_{\textrm{shock}}
\label{vshock1}
,\end{equation}

\noindent 
where $\lambda_{e1}$ is the wavelength of the maximum intensity ($\varphi\simeq 0.90$) of the H$\alpha$ emission and $\lambda_{0}$ its laboratory wavelength.
Since during the first apparition (pulsation phase $\varphi\sim0.91$) the emission line is blueshifted, then $V_{\textrm{shock}}<0$ and the shock wave is 
coming toward the observer. 
However, for practical purpose we use the absolute value of $V_{\textrm{shock}}$ in this paper.
The uncertainties of Eqs.\,\ref{recomb}, \ref{doppler}, and \ref{vshock1} were estimated from the values given in Table\,1 of \citet{fadey04} for the different 
shock velocities and atmospheric densities.
For the considered shock wave model, the validity of Eq.\,\ref{vshock1} is verified for shock velocities from 40\kms to 90\kms, \ie, for Mach numbers 
ranging from 6.2 to 14.

If we apply this equation to the observed \ha line profile in the pulsation phase interval $0.89-0.93$ (Figure\,\ref{ha_film_201704-2}), we obtain an 
estimate of the shock front velocity.
We derive the \ha emission position by Gaussian fitting or first moment method depending on the line shape. 
Velocity uncertainties, which are much larger than measurements uncertainties, are computed with Eq. \ref{vshock1} estimated from the \citet{fadey04} models.
Since the emission line is blueshifted for the first apparition, the shock intensity, \ie, the absolute value of $V_{\textrm{shock}}$, is considered for
the clarity of the analysis.
Most of measurements come from the night of  2017-04-09. 
In order to improve the time resolution, we added an observation from the night of  2017-04-12, which  corresponds to $\varphi=0.925$, assuming that shock velocity 
does not vary too much between consecutive pulsation cycles. 
This assumption is verified because the measured shock velocity at this phase ($+111$\kmsn) is an intermediate value between $+108$\kms and $+158$\kmsn. 
Theses values are listed in Table\,\ref{vshock2017}.

\begin{table}[!ht]
\caption{Doppler velocity of \ha emission line with respect to the stellar rest frame, $|V_{\textrm{shock}}|$, and Mach number value within the pulsation 
phase for the 2017-Apr-08 observation night ($\psi=0.77$) as defined by Eq.\ref{vshock1}. 
Uncertainties are computed with Eq.\ref{vshock1} estimated from \citet{fadey04} models and are much larger than measurement uncertainties. }
\label{vshock2017}
\centering
\begin{footnotesize}
\begin{tabular}{>{$}c<{$} >{$}c<{$} >{$}c<{$} >{$}c<{$}}
\hline\hline
\varphi & V_{e1,\alpha} & |V_{\textrm{shock}}| & {\rm Mach} \\
 & (\textrm{km\,s}^{-1}) & (\textrm{km\,s}^{-1}) & {\rm number} \\ \hline
0.892 & -44.4 \pm 0.6 & 133 \pm 21 & 13 \\
0.911 & -39.3 \pm 0.6 & 118 \pm 19 & 12 \\
0.925 & -37.0 \pm 0.6 & 111 \pm 18 & 11 \\
0.929 & -36.0 \pm 0.6 & 108 \pm 17 & 11 \\
\hline
\end{tabular}
\end{footnotesize}
\end{table}

With an assumed sound velocity of $10$\kmsn, the Mach number is between 11 and 13. 
Thus, with physical conditions present in the atmosphere of RR Lyr, the shock becomes radiative for $M\gtrsim10$. 

The average shock front velocity is around $120$\kms (see Table\,\ref{vshock2017}). 
This value is much greater than the expected maximum velocity of $+40$\kms of the infalling atmosphere (see Fig.\,\ref{na_ha_velocities} near $\varphi=0.82$). 
Our estimate of the shock front velocity is thus consistent with the hypothesis that the intensity of the shock wave must be intense enough to reverse 
the supersonic infall motion of the sodium layer. 

Our observations of the hydrogen emissions were made on the night of April 8, 2017. 
The corresponding Blazhko phase is $\psi=0.77$, \ie, almost halfway between the minimum and the maximum Blazhko. 
This must explain why the observed shock velocity (from $133$ to $108$\kmsn; see Table\,\ref{vshock2017}) is relatively moderate in contrast to those 
which can occur at maximum Blazhko, up to $\simeq170$\kms (Gillet et al., in prep.).

The question is whether the Eq.\,\ref{vshock1} used in this paper can be applied to velocities as high as 110\kms or even higher.
When the shock velocity increases, the hydrogen gas behind the shock front is rapidly fully ionized. 
Moreover, the optical thickness at the H$\alpha$\,central wavelength decreases  when the shock velocity increases. 
Thus, the radiative cooling of the gas increases.
In the limit case of very efficient cooling, the isothermal approximation can be made: the cooling is assumed to be instantaneous. 
Thus, the gas returns to its original pre-shock temperature ($T_1$) just behind the shock front: $T_2=T_1$. 
This  means that the cooling time is very short compared to the dynamic time of the shocked atmosphere.

However, as shown by \citet{fadey04}, for photospheric densities and with a shock front velocity of 90\kmsn, the compression rate within the hydrogen 
recombination zone only reaches a small percent  of the isothermal case. 
Due to the lack of shock models with front velocities up to 200\kmsn, we can only assume that the approximate Eq.\,\ref{vshock1} remains valid in the 
radiation flux-dominated regime.

Finally, we find that in the phase interval $\varphi=0.892-0.929$ the shock velocity decreases all the time. 
Consequently, the observations do not show any acceleration phase of the shock, but only its deceleration phase in the atmosphere. 
The acceleration phase of the shock necessarily occurs between the moment when the compression waves \itv{w1} and \itv{w2} are created by the 
$\kappa-\gamma$ mechanism and the beginning of the deceleration phase of the shock.

\subsection{Phase 3 (0.320): maximum expansion of the sodium layer}

From Table\,3, the photospheric sodium component (blue component) reaches a velocity maximum of $-31.3$\kms at $\varphi=0.991$ after a continuous 
velocity increase from $\varphi=0.911$. 
After $\varphi=0.911$, the velocity  continually decreases. 
This means that, contrary to the \ha layer where the \ha blueshifted absorption is produced, the sodium layer  first accelerates before 
decelerating from $\varphi=0.911$.

At $\varphi=0.320$, the sodium layer stops its ascending motion and starts its slowly infalling motion. 
The sodium layer thus reaches its maximum expansion in the atmosphere at $\varphi=0.320$, while the \ha layer, where the \ha absorption is formed, 
is still ascending. 
The sodium and \ha layers have now an opposite motion and consequently, from this time, an increasing rarefaction zone appears in the atmosphere.   

The generation of rarefaction and compression waves during the development of the pulsation of the atmosphere have already been theoretically 
demonstrated by \cite{hill72}. 
Thus, it appears that the motion of the photospheric layers is rather complicated when a layer reverses its movement in relation to another.

\subsection{Phase 4 (0.455): maximum expansion of the \ha layer}

After the end of the \ha line doubling ($\varphi=0.027$), the \ha profile is in absorption and single. 
Its radial velocity begins to rapidly decrease from $\varphi=0.074$ ($-52.3$\kmsn) to $\varphi=0.455$ where its value reaches $\sim 0$\kms 
(Table\,\ref{vabs2017} and Fig.\,\ref{ha_film_201704-1}).
Consequently, at $\varphi=0.455$, the \ha line-forming region stops its expansion. 
At $\varphi=0.473$, the \ha absorption line becomes redshifted (Fig.\,\ref{ha_film_201704-1}). 
Thus, this part of the atmosphere begins to reverse its movement toward the center of the star.

During the ascension end of the sodium layer, which was driven by the shock now located just above it, the radial velocity of the \ha blueshifted 
component continues to decrease slowly (from -21.1\kms to -18.8\kmsn) as described in Table\,\ref{vabs2017}.
This means that the expansion of the \ha layer located just behind the shock front is not yet complete. 
The \ha profile is now in absorption and single. 
Its radial velocity continues to decrease  to $\sim 0$\kms at $\varphi=0.455$ (Table\,\ref{vabs2017} and Fig.\,\ref{ha_film_201704-1}).
At $\varphi=0.473$, the \ha absorption line becomes redshifted (Fig.\,\ref{ha_film_201704-1}). 
Thus, this part of the atmosphere begins to reverse its movement toward the center of the star. 

\subsection{Phase 5 (0.320-0.455): a two-step infalling motion}

During the phase interval $0.320\lesssim\varphi\lesssim0.455$ the sodium line is redshifted. 
As a result, the lower atmosphere, where the sodium component is produced, begins its infalling movement well before ($\varphi=0.320$) 
that of the atmosphere in which the \ha line is formed ($\varphi=0.473$). 
It is necessary to wait for 15\% of the pulsation period for the hydrogen layer to begin to fall on the star.  

Our observations show that the atmospheric layers do not follow a synchronized ballistic movement even in the lower part of the atmosphere. 
It is the propagation of the radiative (hypersonic) shock wave that causes this strong desynchronization of the pulsation movement. 
If we limit our observational analysis to the movement of the layers of sodium and hydrogen, we have an infalling movement in two successive steps. 
This means that during the phase interval $0.320\lesssim\varphi\lesssim0.455$, the sodium and \ha layers, which are localized behind the shock, 
have opposite movements within the atmosphere.

Finally, it is expected that when a shock of large amplitude (hypersonic) occurs within a pulsating atmosphere, a strong desynchronization of the layer motions takes place in the atmosphere. 
It should be easy to validate this statement using high-resolution spectroscopic observations.

\subsection{Phase 6 ($\sim 0.36$): a strong photospheric compression} 

From our observations presented in a previous paper \citep{gillet2017a}, the \ha third emission line is observed approximately within the phase interval $0.20\lesssim\varphi\lesssim0.40$. 
Due to the variation between pulsation cycles, the duration range of visibility of the third emission may undergo some variations on the order of 5\% to 10\% 
in the pulsation phase. 
In all cases, the presence of a third emission line indicates that a strong warming occurs in the atmosphere due to the intense compression induced by 
the infalling layers.

Due to the new pulsation cycle, as shown by the model, at $\varphi=0.20$ the whole atmosphere located behind the shock wave is expanding.
Only the highest layers of the atmosphere that have not yet been traversed by the shock are still falling back on the star. 
The ballistic motion of these layers was induced by the previous pulsation cycle.

From our observations we find that the \ha line-forming region stops its expansion at $\varphi=0.455$. 
Because the third \ha emission disappears near $\varphi=0.40$, this means that the ascending \ha region contributes all the time to the compression, 
inducing the \ha third emission line. 
On the other hand, this is not always the case of the sodium layer because it begins its infalling movement at $\varphi=0.320$.  

The maximum intensity of the \ha third emission should correspond to the maximum of the compression. 
For our observations made during the night of 2014-09-14 \citep{gillet2017a}, the maximum compression probably occurs around the pulsation phase 
$\varphi=0.357$. 
Thus, the most intense part of the compression occurs after the beginning of the infalling motion of the sodium layer. 

The atmospheric phenomena and the shock wave behavior are summarized in Table\,\ref{layersmoves}.

{\setlength{\tabcolsep}{4.4pt} 
\begin{table*}
\caption{Shock wave, and Na and \ha layer movements in stellar rest frame during a typical pulsation cycle of \rrly, as described in Sect.~4. The given phase values may vary to within a few hundredths because the temporal reproducibility of the cycles is variable.}
\label{layersmoves}
\centering
\begin{footnotesize}
\begin{tabular}{>{$}c<{$}p{0.17\textwidth}p{0.17\textwidth}p{0.47\textwidth}}
\hline\hline
\varphi\ \mathrm{interval} & \centering Na movements & \centering \ha movements & \hspace{0.18\textwidth} Shock wave \\ \hline
0.874-0.90 & Up to about $\varphi=0.90$, the Na layer is receding. When it is traversed by the shock, it reverses its motion and starts an ascending motion in the atmosphere. &
 Highest layer receding at an average velocity near $+46$\kmsn &
 \itv{s1} and \itv{s2} shock waves are initiated below photosphere, but are not observed individually. 
When above the photosphere, the main shock (\itv{s1}+\itv{s2}) is crossing the Na layer. \\
0.90-0.943 & Ascending acceleration from 0 to $-24$\kmsn &
 Highest layer receding at $+46$\kms throughout this phase interval and deepest layer ascending (producing emission in shock wake) &
Deceleration of the main shock from $133$ to about $100$\kmsn. \newline First apparition of \ha emission ($0.892\lesssim\varphi\lesssim0.911$). \\
0.943-1.074 & The Na layer reaches its maximum ascending velocity ($-31.3$\kmsn) at $\varphi=0.991$ &
 Ascending deceleration from $-58$ to $-63$\kmsn. \newline Short line doubling ($0.953<\varphi<1.027$) &
Deceleration of the main shock followed by disappearance (radiative dissipation). The main shock leaves the \ha layer around $\varphi=1.05$. \\ 
1.074-1.302 & Ascending deceleration from $-24$ to $0$\kmsn. The Na layer reaches its maximum radius. &
 Ascending deceleration from $-52$ to $-21$\kmsn &
The main shock has left the \ha formation layer.  \\ 
1.302-1.320 & Maximum expansion of Na layer  & The highest part of the \ha layer is still ascending. &
The main shock has left the atmosphere. 
\newline Apparition of the third \ha emission ($0.20\lesssim\varphi\lesssim0.40$) induced by the compression of the high atmosphere. \\
1.320-1.455 & Receding acceleration from $0$ to $+10$\kmsn. \newline Two-step infalling motion. Strong compression of deep photospheric layers at $\varphi\sim0.36$ &
 Ascending deceleration from $-19$ to $0$\kmsn. Maximum expansion of \ha layer. &
The main shock has left the atmosphere. \\
1.455-1.600 & Receding acceleration from $+10$ to $+24$\kmsn. &
 Receding acceleration from $0$ to $+23$\kmsn &
The main shock has left the atmosphere.  \\ 
1.600-1.874 & Receding acceleration from $+24$ to $+37$\kmsn\ near $\varphi\sim0.82$. Then slight slowdown until  $+32$\kmsn\ at $\varphi=0.874$ just before the arrival of a new main shock. &
 Receding acceleration from $+23$ to $+42$\kmsn. &
A new shock emerges from the photosphere near $\varphi=0.874$. \newline Apparition of second \ha emission ($\varphi\sim0.7$) due to atmospheric compression. \\
\hline
\end{tabular}
\end{footnotesize}
\end{table*}
}

\section{Shock wave dynamics}

\subsection{\ha line components and shock velocity}

At $\varphi=0.943$, the \ha line doubling phenomenon is already present (Fig.\,\ref{ha_film_201704-2}). 
It is observed up to $\varphi=0.027$ when the redshifted component disappears. 
The velocity of the latter remains virtually constant: +50\,\kms from $\varphi=0.855$ to $\varphi=0.027$, hence just before the   emission, 
during its presence,  after its disappearance, and until the end of the line doubling phenomenon. 
In addition, the velocity of the blueshifted \ha component decreases all the time from -58.5\,\kms ($\varphi=0.943$) to -52.3\,\kms ($\varphi=0.074$). 
This means that the motion of the \ha layer located just behind the shock front undergoes a continuous but modest deceleration (the velocity decreases only 
by 11\% over 13\% of the pulsation cycle).

The estimation of the velocity of the shock from the radial velocity of the blueshifted \ha absorption component of the line doubling profile is not obvious. 
This component is not formed in the radiative wake of the shock wave, \ie, just behind the shock front; 
on the contrary, it is formed far beyond the shock, within atmospheric layers that undergo complex movements due to the possible presence of rarefaction waves. 
Under these conditions, only a complete model of the dynamics of the atmospheric layers, taking precisely into account the radiative dissipation of the 
shock and the dynamical interaction of layers, should allow us  to obtain a pertinent estimate of the shock front velocity. 
Unfortunately, today such a model is not available.

\subsection{Shock wave velocity evolution}

All the observations presented in this paper were made between April 8 and 18, 2017, as described in Sect.\,2.
The Blazhko phase $\psi$ was approximately $0.77$ at the beginning of the observations and 0.03 at the end, \ie, at a Blazhko maximum. 
The \ha emission is clearly visible at the pulsation phases $\varphi=0.892$ and $\varphi=0.911$. 
These two spectra were obtained during the same pulsation cycle, which occurred on the night of April 8, 2017. 
Using Eq.\,\ref{vshock1}, we obtain a shock velocity of 133 and 118\,\kms, respectively, for these two phases (see Table\,\ref{vshock2017}). 

The maximum velocity of the shock is variable from one Blazhko maximum to another. 
For example, by August 13, 2014, the \ha profile was observed \citep{gillet2017a} very close to a Blazhko maximum ($\psi=0.04$). 
It shows that the shock velocity decreases rapidly from 156\;km\,s$^{\rm -1}$ at $\varphi=0.903$ to 101\;km\,s$^{\rm -1}$ at $\varphi=0.941$. 
From H$\alpha$ profiles observed near $\psi\cong0.10$ by \citet{GFL2013}, \cite{gillet14} estimated the shock velocity from 132\;km\,s$^{\rm -1}$ 
at $\varphi=0.902$ to 114\;km\,s$^{\rm -1}$ at $\varphi=0.927$.
 
Finally, these estimates clearly show that the values of the shock velocity as well as the rate of deceleration are highly variable along the Blazhko cycle. 
Thus, it would be interesting to carry out a high-quality observational study to improve our knowledge on the variation in the shock velocities occurring 
in RR\,Lyr during a Blazhko cycle and in different Blazhko cycles.

In Fig.\,\ref{ha_film_201704-1}, the \ha third emission is clearly visible from phase $\varphi=0.202$ to $\varphi=0.382$.
As discussed above, during this phase interval, the shock wave velocity is probably less than 50\kmsn.
Below this velocity, the shock is no longer energetic enough to produce an emission component in its wake in the \ha profile:
the shock is no longer radiative.

\begin{figure}
  \centering
  \includegraphics[width=0.96\hsize]{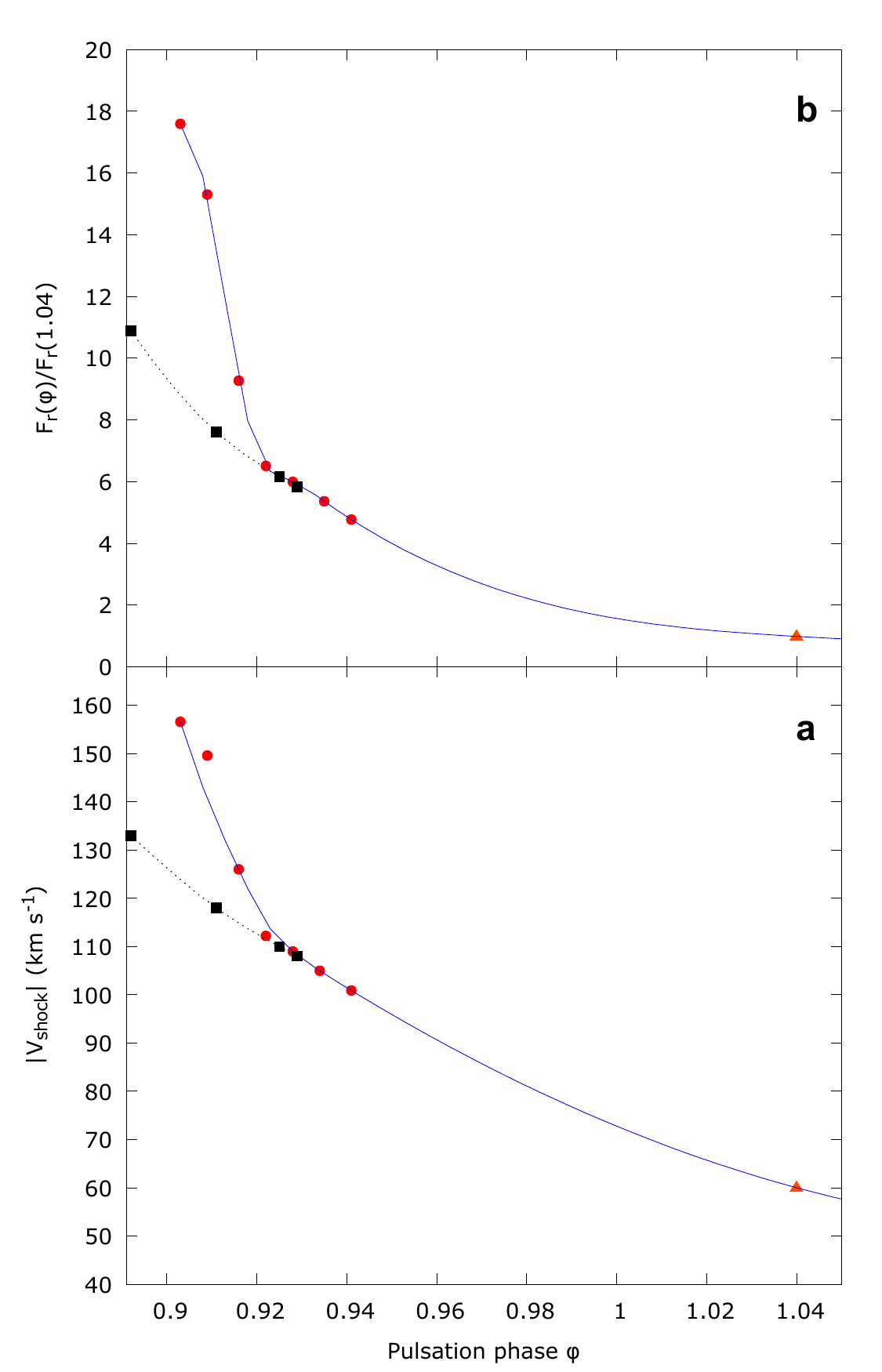}
  \caption{Phased kinetic and radiative flux measurements with trend curve.
\newline Measurements points are the most probable values of $|V_{\textrm{shock}}|$ and $F_r(\varphi)/F_r(1.04)$. Associated uncertainties are computed in Tables\,\ref{vmes} and \ref{vshock2017}. 
Red dots are measured values from 2014-Aug-13 while the red triangle represents the night of 2013-Oct-12. 
Velocities from April 2017 are black squares with the corresponding trend curve (grey dotted line).
\newline \textbf{Panel a:} Evolution of $|V_{\textrm{shock}}|$ plotted with a cubic spline interpolation curve. 
In the phase interval $\varphi=0.90-0.92$, $V_{\textrm{shock}}$ decreases much more than in the phase interval $\varphi=0.92-1.04$. 
Thus, for 40\% of the period, the shock front velocity decreases by a factor of more than three.
\newline \textbf{Panel b:} Evolution of maximum values of $F_r(\varphi)/F_r(1.04)$ ratio plotted with a cubic spline interpolation curve. 
In the phase interval $\varphi=0.90-0.92$, the $F_r$ ratio decreases much more than in the phase interval $\varphi=0.92-1.04$. 
The flow of radiative losses increases rapidly with the speed of the shock front.
}
  \label{trio_vae}
\end{figure}

\medskip
For the 2017 run, the time resolution did not allow us to precisely see how $|V_{\textrm{shock}}|$ evolves during a pulsation cycle. 
We thus used a previous set of spectra obtained during 2014-08-13 and 2013-10-12 with exposure times of 300\,s to enable such a study. The 
$|V_{\textrm{shock}}|$ values are given in Table\,\ref{vmes}.
Hence, from the 2014-2013 data set, it was possible to compute a cubic spline interpolation within a 0.05 phase resolution.
All measurements are listed in Table\,\ref{vmes}. 
Phased kinetic and radiative flux measurements are plotted with trend curves in Fig.\,\ref{trio_vae} to show the global evolution.

{\setlength{\tabcolsep}{5pt} 
\begin{table}[!ht]
\caption{$|V_{\textrm{shock}}|$ and $F_r(\varphi)/F_r(1.04)$ within the pulsation phase for 2014-Aug-13 and 2013-Oct-12 observations. 
The shock velocity was calculated from the emission of \ha for all phases except for the one at 1.040 (\ion{He}{i}).
The uncertainties, which are much larger than measurements uncertainties, are computed using Eq.\ref{vshock1} from the \citet{fadey04} models .
}
\label{vmes}
\centering
\begin{footnotesize}
\begin{tabular}{>{$}c<{$} >{$}c<{$} >{$}c<{$} c c}
\hline\hline
\varphi & |V_{\textrm{shock}}| \; (\textrm{km\,s}^{-1}) & F_r(\varphi)/F_r(1.04) & Date & $\psi$ \\ \hline
0.903 & 156 \pm 25 & 17.6 \pm 11.8 & 2014-08-13 & 0.04 \\
0.909 & 149 \pm 23 & 15.3 \pm 10.3 & 2014-08-13 & 0.04 \\
0.916 & 126 \pm 20 & 9.3 \pm 6.2 & 2014-08-13 & 0.04 \\
0.922 & 112 \pm 18 & 6.5 \pm 4.4 & 2014-08-13 & 0.04 \\
0.928 & 109 \pm 17 & 6.0 \pm 4.0 & 2014-08-13 & 0.04 \\
0.935 & 105 \pm 17 & 5.4 \pm 3.6 & 2014-08-13 & 0.04 \\
0.941 & 101 \pm 16 & 4.8 \pm 3.2 & 2014-08-13 & 0.04 \\
1.040 &  60 \pm 9  & 1.0 \pm 0.7 & 2013-10-12 & 0.13 \\
\hline
\end{tabular}
\end{footnotesize}
\end{table}
}

\medskip

In order to get an idea of the rate of decay of the shock velocity during its elevation in the atmosphere, we have the shock velocity estimate 
at $\psi=0.04$ (August 13, 2014) from $\varphi=0.903$ to $\varphi=0.941$ in addition to those at $\varphi=1.04$ (October 12, 2013, $\psi=0.13$). 
These values are listed in Table\,\ref{vmes} and displayed in Fig.\,\ref{trio_vae}a. 
These shock velocities  come from two different Blazhko cycles, but they are all close to a Blazhko maximum. 
They cover the phase interval from 0.90 to 1.04, \ie,  14\% of the pulsation period. 
They correspond to the appearance in the atmosphere of the main shock ($\varphi=0.90$) until it disappears ($\varphi>1.04$) as a radiative shock, \ie, 
producing the \ha and \ion{He}{i} emissions within the shock wake.
Although the shock velocities  come from two different Blazhko cycles, we clearly identify a decrease in the shock velocity of a factor of about three. 
Thus, during this strong deceleration phase, the shock propagates over a distance of about two solar radii, \ie, over 30\% of the stellar radius $R_\star$. 
When the P Cygni profile of the D3 helium line is observed at phase $\varphi=1.04$, the atmospheric extension of 16\% is not very high, but  high 
enough to already observe the characteristic P Cygni shape.
Thus, we confirm that the shock is well detached from the photospheric disk at the pulsation phase $\varphi=1.04$.

Uncertainties are computed using Eq.\,\ref{vshock1}. 
They represent the dispersion of the resulting values of the models calculated by \cite{fadey04} for different shock velocities and unperturbed densities 
used as parameters in their Table\,1. 

As discussed in Section\,5.1, from our observations of the sodium and \ha line profiles, it is not possible to distinguish independently the 
shock waves s1 and s2. 
We only observe the resulting shock s4+s3+s3'+s1+s2, called the main shock, obtained from the successive merging of s1, s2 and the secondary shocks s4, s3, and s3'.
In their model RR41, \cite{fokin97} show the fusion of the two shocks s1 and s2. 
Moreover, they show that the amplitude of the shock s4+s3+s3'+s1 decreases rapidly first after merging s1 and again after merging s2 (see their Fig.\,6). 
This weakening of the shock amplitude is consistent with our observations of the decreasing shock velocity when it rises in the atmosphere, as shown in Fig.\,\ref{trio_vae}. 
Thus, in the context of the physical and dynamic conditions taking place in the RR\,Lyr atmosphere, the shocks do not seem to present an accelerating phase above the photosphere.

\subsection{Shock weakening regimes}

During its propagation in the atmosphere, the shock is continuously decelerated (Fig.~\ref{trio_vae}). 
It loses energy in two forms. 
One form is density dilution ($1/r^{2}$) as its radius $r$ increases. 
The other involves radiative losses as soon as it reaches a hypersonic speed, \ie, when its Mach number is  greater than approximately 5. 
Using the Rankine-Hugoniot equation for energy and assuming that the shock is strong (until the isothermal limit), the maximum value of the radiative 
flux $F_r$ produced by the shock is (see Appendix) 

\begin{equation}
F_r\leq \frac{1}{2}{\rho_{1} V_{\textrm{shock}}^{3}}\,,
\label{eq:frad}
\end{equation}
 
\noindent 
where $\rho_{1}$ is the density of the unperturbed gas in front of the shock and $V_{\textrm{shock}}$ the shock front velocity. 
Thus, the flow of radiative losses increases rapidly with the shock front speed (to the third power). 
In the context of a realistic model, \cite{fadey01} estimated the irreversible loss occurring in the shock wave energy. 
Comparing the radiation flux emerging from the boundary of the shock layer with the total energy flux, they showed that the radiative losses 
increase very rapidly with increasing shock front velocity. 
They showed that 70\% or more of the shock energy is lost in radiative form as soon as the shock front velocity is greater than 50\,\kms for physical 
conditions occurring in the atmosphere of pulsating stars with unperturbed gas of temperature between $T_1=3\,000$\,K and $T_1=6\,000$\,K with a 
density $\rho_{1}=10^{-10}$\;g\,cm$^{\rm -3}$. 

Consequently, applying the equation above, the shock would lose, in first approximation, up to 18 times more radiation energy at $\varphi = 0.903$, 
15 times at $\varphi=0.909$, and only 6 times at $\varphi=0.928$ with respect to $\varphi = 1.040$ (see Fig.\,\ref{trio_vae}b). 
Accordingly, the shock wave will probably  lose most of its energy before the maximum brightness of the star.
In 2017, when the maximum velocity of the shock was smaller (133\kms instead of 156\kmsn), the radiative losses only represent 11 times those of the adiabatic regime. 
For the 2017 data we computed the radiative flux ratio with the 2014 value of $F_r(1.04)$, \ie, $V_{\mathrm{shock}}(1.04)$, assuming that 
$V_{\mathrm{shock}}$ converges asymptotically to the same value as 2014, as shown in Fig.\,\ref{trio_vae}.

In 2014, the deceleration is not uniform between the pulsation phases 0.90 and 1.04 (Fig.\,\ref{trio_vae}a). 
A two-step deceleration phase clearly appears.
First, during a very short interval of the pulsation cycle (5\% from $\varphi=0.90$ to $\varphi=0.95$), the shock velocity decreases by 
one-third of its value. 
This rate of fast decrease in the shock velocity occurs when the shock has a very large radiative loss, close to that of an isothermal shock.
Then, up to $\varphi=1.04$, but slower, the shock velocity decreases by about a factor of $40\%$ over $10\%$ of the period. 
During this second phase, the radiative losses of the shock become secondary with respect to those induced by the geometric dilution due to the shock propagation. 
These two phases are clearly visible in Fig.\,\ref{trio_vae}. 

In 2017, the maximum shock velocity did not reach a value (133\kmsn) as large as in 2014 
(156\kmsn), \ie, a Mach number of 13 instead of 16. 
This difference seems sufficient to induce the transition between the two hydrodynamic regimes: the isothermal one, in which the radiative losses dominate, 
and the adiabatic one, in which the losses due to the geometric dilution are dominant. 

Finally, when the intensity of the main shock wave is strong enough (isothermal phase), most of the shock wave energy is dissipated by radiative processes at the photospheric level, \ie, in the deep atmosphere. 
The remainder of the shock wave energy is dissipated by dilution in the high atmosphere when the radiative nature of the shock is no longer dominant (adiabatic phase).

\subsection{Turbulent state of the atmosphere}

\cite{gillet1998} discussed the turbulence amplification in the atmosphere of a radially pulsating star, showing that it is   due to the global 
compression of the atmosphere during the pulsation, and that it is also affected by strong shock waves propagating from the deep atmosphere. 
As already noticed by \cite{kolenberg2010} and \cite{fossati2014}, among others, by definition the microturbulence velocity $V_{mic}$ characterizes a 
difference between observed and theoretical atmospheric parameters affecting the FWHM of an absorption spectral line:

\begin{equation}
V_{mic}\equiv\sqrt{FWHM^{2}_{obs}-FWHM^{2}_{model}}.
\label{eq:Vmic}
\end{equation}

\noindent
Therefore, it represents only the upper limit of atmospheric turbulence.
Indeed, all line broadening effects such as velocity gradients or non-LTE effects are not always taken into account in atmosphere models.

A time variation in the microturbulence velocity $V_{mic}$ in the atmosphere of RR Lyrae has been estimated from a comparison of the observed and 
theoretical full width at half maximum (FWHM) curves of the \ion{Fe}{ii} $4923.921$\,\AA\ absorption line by \cite{fokin1999}. 
This velocity $V_{mic}$ reflects motions on scales smaller than the line-forming region. 
\cite{fokin1999} found that turbulence in this region varies from 2 to 7\kmsn, over a large phase interval 
($0.5\lesssim\varphi\lesssim1.0$). 
The maximum (7\kmsn) occurs at the pulsation phase $\varphi=0.7$ (during the light bump) and a secondary peak (3.5\kmsn) occurs at $\varphi=0.2$. 
This prominent peak of $V_{mic}$ starts to appear when the two secondary shocks s4 and s3, as noted by \cite{fokin97}, merge at $\varphi=0.52$ when $V_{mic}$ is at its lowest level (just after the phase of the maximum radius).
The physical origin of the shock s4 is due to the accumulation of several weak compression waves at the sonic point during the beginning of the 
atmospheric compression, while s3 comes from the stop of the hydrogen recombination front near the phase of maximum expansion. 
In the case of RR\,Lyr, the peak maximum of the microturbulence velocity at 7\kms takes place when s4+s3 merge with the shock s3' assumed to be caused 
by the 1H-mode \citep{fokin97}. 
At this stage, the overall velocity of the shock s4+s3+s3' is at a maximum, but still modest (55\kmsn), although supersonic (Mach number $M\gtrsim5$). 
Then the velocity of this shock slowly decreases almost to 40\kmsn (\citealt{fokin97}, their Fig.\,6). 
As a consequence the microturbulence velocity decreases until its rapid and abrupt rise induced this time by the arrival of the strong shocks s2 and s1 
near $\varphi=0.91$, which have their origin in the $\kappa$-$\gamma$ mechanism associated with the H- and He-ionization zones. 
However, the effect of these two last shocks on the microturbulence velocity remains modest because they merge with s4+s3+s3' in the upper part of the line-forming region  of the \ion{Fe}{II} line. 
 
Recently, as part of an in-depth spectroscopic analysis of the Blazhko star RR Lyrae, \cite{kolenberg2010} deduced a microturbulence velocity curve from the FWHM of the 
\ion{Fe}{ii} line at $4508.288$\,\AA. 
In a complementary work, \cite{fossati2014} showed this curve (green solid line in their Fig.\,6); it presents an asymmetrical peak starting 
abruptly at $\varphi=0.9$, with a maximum near $\varphi=0.95$, and then a slow decrease until $\varphi=0.2$. 
It is centered just before the middle of the rising branch of the V-light curve.
The minimum microturbulent velocity occurs at the maximum radius during the most stable pulsation phase, called the quiet phase, near $\varphi=0.33$. 
The value of the microturbulence velocity obtained by these authors is much higher (between 15 and 35\kmsn) compared to the values between 2 and 7\kms found by \cite{fokin1999}. 
This may originate, at least in part, from the use of a static atmosphere model that is not able to take into account the strong velocity gradients present in the star.
Indeed, such a velocity gradient is an important source of line broadening within a pulsating atmosphere (see \citealt{FGB1996}).
From the range of line formation depths covered by the measured lines, \cite{fossati2014} estimate the $V_{mic}$ profiles in the corresponding optical 
depth range $\log\tau_{Ross}=-4$ to $\log\tau_{Ross}=0$, \ie, within the middle part of the atmosphere. 
The most striking fact is that the turbulent velocity continually decreases while the depth in the atmosphere increases. 
This is true for all pulsation phases---especially during the quiet phase and during the maximum compression induced by the global atmospheric pulsation---suggesting that the large-scale motions have no effect on the microturbulent velocity, , a disturbing result.
Between the bottom and the top of the line-forming region, the microturbulence velocity varies from 3 to 20\kms except during the peak where the microturbulent velocity may be as high as 50\kmsn, \ie, largely supersonic.
In addition, \cite{fossati2014} also point out that the microturbulence velocity of the deepest atmospheric layer increases, while that of the layer just above remains unchanged, also a surprising behavior. 

As shown by \cite{gillet1998}, the turbulence amplification during a compression of the atmosphere without the presence of waves results from the variation of the density within the line-forming region. 
Because in radially pulsating stars, such as RR Lyrae stars, the atmospheric curvature is very small, it is possible to assume that the atmospheric distortion is not locally spherical but almost radial.
Thus, compression must induce a preferential amplification direction. 
\cite{gillet1998} estimated that the global atmospheric compression can be almost considered as an adiabatic homogeneous axial compression. 
Consequently, in the framework of a rapid distortion, the solenodial RDT model of \cite{JCB1993} appears to give the better quantitative prediction. 
As an application, for the radially pulsating star $\delta$\,Cephei, the predicted turbulence amplification induced by the global atmospheric 
compression is consistent with the solenodial RDT \citep{gillet1998}.

In the case of the presence of strong shocks as in the case of RR\,Lyrae stars, \ie, when the compression rate becomes  higher than 2 
(Mach number $\gtrsim 2$), radiative effects take place and adiabatic turbulence amplification theories break down. 
In this case, these theories give a considerable overestimation of the amplification \citep{gillet1998}.
Consequently, in the limiting case of very strong shocks (isothermal shocks), the compression ratio $\eta$ caused by the shock is $\eta\simeq\,M^{2}$. 
Basically, effects induced by radiative terms in conservation equations are certainly at the origin of the observed limit of the turbulence amplification. 
Nevertheless, the relevant theory does not yet exist and a new theoretical approach is required to take into account the radiative field within 
hypersonic shocks.

In this paper, we have established that the movements in the atmosphere are complex because the motion of atmospheric layers are not synchronous 
and are disturbed by the passage of strong shock waves. 
It is not possible to observe the steps of formation and acceleration of the main shock. 
Only its long deceleration phase is observable. 
During slightly less than 15\% of the pulsation cycle ($0.90\leqq\varphi\leqq1.04$), the main shock velocity decreases from 150 to 60\kmsn, 
\ie, from a Mach number of 15 to 6. 
This strong deceleration occurs in only 2\;hours, a rather short duration compared to the 13.6\;h period. 
The radiative flux emitted by the shock then drops by a factor of 18 (Fig.\;\ref{trio_vae}.b), demonstrating that its energy is essentially dissipated in radiative form. 
All atmospheric layers do not reach their maximum expansion at the same time (a 2-hour difference between sodium and H$\alpha$ as shown in Fig.\;\ref{na_ha_velocities}), generating rarefaction waves during the pulsation from $\varphi=0.30$ to $\varphi=0.50$ (see Table\,5). 
Then a violent atmospheric compression occurs in the lower part of the atmosphere at 
$\varphi=0.36$ due to the rapid fall of the upper layers of the atmosphere.

Therefore, in the context of the extremely complex atmospheric pulsation of RR\,Lyr, it is not obvious to validate the use of successive static atmosphere models, assuming LTE and plane-parallel geometry, to interpret the different phenomena involved in this highly dynamic atmosphere.
As a result, it seems difficult to get a good determination of the effective temperature through synthetic spectra fitting of the observed H$\gamma$ line, and of the surface gravity, which is obtained by imposing the ionization equilibrium for iron.
Moreover, while imposing the equilibrium between the iron line abundance and equivalent widths to determine
the microturbulent velocity, \cite{fossati2014} introduced the need of a depth-dependent  $V_{mic}$.
Thus, in the context of a very disturbed atmosphere revealed by the observations reported in this article, it would be interesting to confirm the result of \cite{fossati2014} regarding the depth-dependent of $V_{mic}$.

For $\delta$\,Cephei, the maximum compression rate within the \ion{Fe}{II} line-forming region induced by the global compression of the atmosphere near $\varphi=0.85$ is of the same order of magnitude as that caused by the main shock \citep{gillet1998}.
Because the maximum amplitude of the latter in RR\,Lyr is about 5 times larger than that of $\delta$\,Cephei, we must expect that the dominant 
amplification phenomenon could be the main shock and not the global atmospheric compression.
Moreover, as shown at the beginning of this section, the intensity of the main shock of RR\,Lyr is extremely large ($M\gtrsim15$) only in the lowest 
part of the atmosphere because the shock velocity decreases very rapidly when it rises in the atmosphere, due to the intense radiative loss occurring 
in the shock wake. 
Consequently, the greatest microturbulence velocities should be observed essentially in the lower part of the atmosphere.
Furthermore, as shown by \cite{gillet1998}, the presence of several secondary shocks in the atmosphere, as expected by pulsation models \citep{fokin97}, 
can also be strong enough in the \ion{Fe}{II} line-forming  region to contribute to an amplification of the microturbulence velocity on the order 
of that induced by the global compression of the atmosphere. 
This is consistent with the result obtained by \cite{fokin1999} because the wide maximum of microturbulence velocity spanning half a pulsation period 
occurs when secondary shocks s3 and s3' are crossing the line-forming  region. 
Also, the third H$\alpha$ emission observed near the pulsation phase 0.3 \citep{gillet2017a} is interpreted as the consequence of a large 
atmospheric compression which provokes a strong and rapid increase of density and thus an appreciable amplification of the microturbulence velocity. 
This atmospheric compression could explain the secondary peak (3.5\kmsn) of microturbulence velocity over the phase interval $0.10\leqq\varphi\leqq0.35$. 

Finally, in order to confirm  that the microturbulence velocity increases with height in the atmosphere, as found by \cite{fossati2014}, 
it would be necessary to perform for RR\,Lyr a computation using a convective pulsation model including an extended atmosphere and allowing the generation of shock waves of large amplitude.

\section{Conclusion}

The objective of this paper was to determine from spectral observations the main dynamic phenomena occurring in the atmosphere of \rrly during a typical pulsation cycle. 
As the intensity of the pulsation cycles varies greatly over time, in particular because of the Blazhko effect, we considered a mean amplitude cycle, \ie, 
observations with a Blazhko phase close to $0.75$. 
The atmospheric phenomena that can occur during the cycles of extreme amplitude will be examined in a second paper.

This observational study was based on a series of 79 spectra sampled approximately every 15\,min on the pulsation cycle. 
Thus, we are close to a temporal resolution of two hundredths (period $\sim$13\,h\,36\,min). 
This resolution seems quite sufficient to follow all the developments occurring in spectral line profiles such as \ha and Na\,D. 
Moreover, although the  resolving power used is relatively modest ($R=11\,000$), all the potential spectral information that could exist in the 
profiles seems to be highlighted. 
All the spectra of this series were obtained between  7 and  18 April 2017; therefore, during the same Blazhko cycle. 
Observations were started at the Blazhko phase $\psi=0.75$ and were stopped at maximum Blazhko ($\psi=0.03$). 
Since the most intense hydrogen emission occurs at the beginning of the observation series, the intensity of the main shock corresponds to the 
average pulsation cycle ($\psi\sim0.77$).

It should be noted that these observations were made with a modest telescope (35\,cm). 
As \rrly is the brightest \rrlyr star ($7<V<8$), an effective exposure time of 900\,s allows a S/N of around 100 per pixel depending on 
weather conditions. 
As the feasibility conditions of this spectroscopic study were satisfying for this star, its extension to other \rrlyr stars requires at least a one-meter class telescope.

\bigskip

The observed sodium and \ha profiles presented in this paper did not allow us to resolve the shocks \itv{s1} and \itv{s2} individually. 
Only the main shock, the sum of these two shocks induced by the $\kappa$-mechanisms in the He and H-ionization subphotospheric layers, is confirmed. 
It is possible that observations with higher time resolution (exposure time less than 5\,min), high signal-to-noise ratio ($S/N\gtrsim100$) 
and high spectral resolution ($R\gtrsim50\,000$) may allow the detection of these two shocks separately. 
Indeed, the observation of the shock \itv{s2}, which is initiated deep below the photosphere, is probably impossible due to the excessive opacity of the gas. 
To observe \itv{s2}, it would have to have existed above the photosphere before it merged with \itv{s1}, 
but this does not seem to be the case from our observations.

One of the most striking results of this observational study is that we do not observe the acceleration phase of the main shock as soon as the \ha emission 
produced in the shock wake becomes visible. 
Consequently, we directly observe the deceleration of the wave. 
The deceleration is faster when the shock is radiative than after when dilution phenomena become dominant.
In the deep atmosphere (for 0.903<$\varphi<0.941$), the shock is radiative when its Mach number is larger than 10, whereas when in the high atmosphere 
($\varphi=1.040$) a Mach number around 6 is enough. 

The sodium layer reaches its maximum expansion at $\varphi=0.320$, while for the layer in which the \ha line core is formed, it is much later, at $\varphi=0.455$. 
Thus, a rarefaction wave appears in the atmosphere. 
This phenomenon is clearly predicted by the models of \cite{fokin97}.

The so-called third \ha emission is observed around $\varphi\sim0.36$. 
It occurs when the ascending \ha layer and the highest infalling atmospheric layers of the previous pulsating cycle compress the gas located between them. 
At this pulsation phase ($\varphi\sim0.36$) the sodium layer, located in the deep atmosphere,  already has an infalling movement, which means that the sodium does 
not participate in the compression inducing the third \ha emission.

As previously shown by \cite{gillet1998}, the turbulence amplification of the atmospheric gas is mainly due to both the global compression of the 
atmosphere during its pulsation and to strong shock waves propagating through the atmosphere. 
In the case of RR\,Lyr, because amplitudes of shocks occurring within the atmosphere are large, the turbulence amplification is mostly caused by shocks, 
while that provoked by the global compression of the atmosphere seems secondary. 
This is the reason why the main peak of microturbulence velocity is very wide for RR\,Lyr conversely to the case of $\delta$\,Cephei where it is quite narrow. 
Furthermore, because the main shock is always observed with a rapidly decreasing velocity during its propagation, the consecutive induced turbulence amplification 
should be highest in the lower part of the line-forming region of the \ion{Fe}{II} line. 
It would be interesting to establish whether the microturbulence velocity increases or decreases
with height in the atmosphere; this might be accomplished with a pulsation model including convection and an extended atmosphere and allowing the generation of shock waves of large amplitude.

\smallskip

In either case, in the future it would be constructive to know the variations in the shock intensity present in the atmosphere of \rrlyn, 
as well as those occurring in the various line profiles induced by atmospheric dynamics. 
However, this study requires a large observational survey because it is essential to integrate the effects due to the Blazhko phenomenon on atmospheric dynamics. 
This will be the goal of the second article in this series.

Finally, the data and work presented in this paper demonstrate further the increasing role of the amateur spectroscopy community in stellar surveys.


\bigskip
\begin{acknowledgements}
We thank \textit{Lux Stellarum} and the French OHP-CNRS/PYTHEAS for their support.
We also thank D. Boureille for his help on smart small tools. 
The present study has used the SIMBAD database operated at the Centre de Donn\'ees Astronomiques (Strasbourg, France) and the GEOS RR\,Lyr database hosted by IRAP (OMP-UPS, Toulouse, France), created by J.F. Le Borgne.
Spectroscopic data have been analyzed with the SpcAudace software written by B. Mauclaire (ARAS group, France). 
We especially thank Christian Féghali for his very careful reading of the manuscript and for his pertinent remarks. We appreciate the helpful comments from the reviewer and editor.
We gratefully acknowledge Helenka Kinnan for her very careful reading of the final version of this paper.

\end{acknowledgements}


\bibliographystyle{astroads}
\bibliography{rrlyr_biblio}

\begin{appendix}
\section{The shock model of \cite{fokin97}}

\cite{fokin97} computed more than 20 nonlinear, nonadiabatic, purely radiative pulsation models of the \rrlyrn\  prototype, \rrlyn. 
However, since \rrly is located very close to the red edge of the instability strip, convection probably slightly affects the ionization zones of 
hydrogen and helium at the origin of the pulsation.
These models go through the subphotospheric layers up to an extended atmosphere.

The shock waves are implicitly calculated with the Von Neumann-Richtmyer artificial viscosity that mimics the viscous dissipation of the 
kinetic energy of the shock. 
The consequence of this artificial viscosity is to widen the shock wake. 
Thus, it is not possible with this kind of numerical approach to resolve the fine structure of the shock because the spatial resolution is too weak 
(about $10^{3}$\,km), or even to resolve the largest zone of the wake where photorecombinations occur (typically over a few\,km). 
As part of their numerical approach,\, \cite{fokin97} note that the differentiation between a wave and a shock wave is not always obvious through 
the numerical model. 
They used the word ``shock'' when the density jump exceeds 5 over four consecutive mass zones.

The effect of excessive damping of the pulsation and waves due to the artificial viscosity have been tested by \cite{fokin97}. 
It appears that their models are not very sensitive to the  damping and cutoff parameters used. 
Consequently, the characteristics of the shocks and their associated hydrodynamics phenomena are stable and are present from one cycle to another.

During each pulsation cycle, four main shocks, with a well-identified physical origin, are produced in the outer part of the stellar envelope. 
In the case of \rrlyn, a fifth shock (shock \itv{s3'}) is also highlighted; however,   its physical origin has not yet been clearly  identified. 
Contrary to the static regime, the extension of the atmosphere increases by several times during the pulsation. 
It is interesting to note that the four main shocks are stronger than in classical Cepheids, but their general features, especially their physical origin, are qualitatively similar.

The two strongest shocks are the consequence of the recombination of the ionization zones of hydrogen and helium in the subphotospheric region. 
\cite{fokin97} call theses shocks \itv{s1} and \itv{s2}, respectively. 

The scheme of phenomena predicted by \cite{fokin97} model are summarized in Table\,\ref{toc1997model}. 
Their models (including RR41) limit the description of the atmosphere to an external density to $10^{-14}$\;g\,cm$^{\rm -3}$. 
Thus, the maximum atmospheric height above the photosphere only reaches $0.22R_{ph}$ near the phase of the maximum expansion ($\varphi=0.55$).

{\setlength{\tabcolsep}{4.4pt} 
\begin{table}[!ht]
\caption{Scheme of phenomena from the  \cite{fokin97} model (RR41) during a pulsation cycle.}
\label{toc1997model}
\centering
\begin{footnotesize}
\begin{tabular}{>{$}c<{$}p{0.37\textwidth}}
\hline\hline
\varphi\ \mathrm{interval} & \hspace{0.15\textwidth} Phenomenon \\ \hline
0.75-0.80 & The $\kappa-\gamma$ mechanism associated with He-ionization and H-ionization zones provokes local overpressure, which transforms into 
outward progressive waves \itv{w1} and \itv{w2}. \\
0.80/0.93 & \itv{w1} and \itv{w2}  rapidly become shocks \itv{s1} and \itv{s2}, which emerge from the photosphere. \\
0.96 & \itv{s1} merges with secondary shocks. \\
1.01 & \itv{s2} merges with \itv{s1}; apparition of the main shock. \\
1.05 & The main shock leaves the described atmosphere. \\
1.35 & Maximum expansion of the photosphere. \\
1.60 & Maximum expansion of the highest atmospheric layers. \\
\hline
\end{tabular}
\end{footnotesize}
\end{table}
}

\section{Maximum radiative flux produced by the shock}

This section describes which hypotheses are applied on the Rankine-Hugoniot equation system to obtain an upper limit on the radiation flux relation 
(Eq. \ref{eq:frad}) used in this paper.

\medskip

Considering the total energy Rankine-Hugoniot relation, the radiation occurs through three different terms: 
the radiative pressure $p_{r}$, the specific radiative energy $e_{r}$, and the radiative flux $F_{r}$: 

\begin{equation}
\frac{1}{2}u^{2}+e_{g}+e_{r}+\frac{p_{g}}{\rho}+\frac{p_{r}}{\rho}-\frac{F_{r}}{{\rho}{u}}={\rm constant}
\label{eq:rher}
\end{equation}

\noindent In general, the exact value of the three radiative terms is difficult to determine, especially when the gas opacity is strongly frequency dependent. 
The first three angular moments of the specific intensity $I_{\mu\nu}$ of the radiation field are (\cite{miha84}, Sects. 6.4-6.6)

\begin{equation}
e_{r}\equiv\frac{1}{c}\int_{0}^{\infty}\,d\nu\int_{-1}^{+1}I_{\mu\nu}\,d\mu\qquad(\textrm{radiative energy}),
\label{eq:der}
\end{equation}

\begin{equation}
F_{r}\equiv\frac{1}{c}\int_{0}^{\infty}\,d\nu\int_{-1}^{+1}\mu\,I_{\mu\nu}\,d\mu\qquad(\textrm{radiative flux}),
\label{eq:dFr}
\end{equation}

\noindent and

\begin{equation}
p_{r}\equiv\frac{1}{c}\int_{0}^{\infty}\,d\nu\int_{-1}^{+1}\mu^{2}I_{\mu\nu}\,d\mu\qquad(\textrm{radiative pressure}),
\label{eq:dpr}
\end{equation}

\noindent where $\mu\equiv\cos\theta$ is the cosine of the polar angle $\theta$ between the direction of the considered radiation beam and the normal to the 
emitting surface. 
Nevertheless, when the fluid is close to the thermodynamical equilibrium, \ie, when it is optically thick, then the evaluation of the  angular moments $I_{\mu\nu}$  is 
straightforward. 
In this case (see, e.g., \cite{samp65}, Sect. 5.1 or \cite{miha84} Sects. 6.8 and 6.9) the specific radiative energy is

\begin{equation}
e_{r}=e^{*}_{r}\equiv\frac{a_{r}T^{4}}{\rho}
\label{eq:er*}
\end{equation}

\noindent and

\begin{equation}
p_{r}=p^{*}_{r}\equiv\frac{a_{r}T^{4}}{3}=\frac{\rho\,e^{*}_{r}}{3}\,,
\label{eq:pr*}
\end{equation}

\noindent where $a_{r}\equiv4\sigma_{r}/c=7.56\,10^{-15}$\,erg\,cm$^{-3}$\,K$^{-4}$ is the radiation constant and 
$\sigma_{r}=5.67\,10^{-5}$\,erg\,s$^{-1}$\,cm$^{-2}$\,K$^{-4}$ is the Stefan-Boltzmann constant. 
At the strictly thermodynamic equilibrium limit, the radiation field is isotropic. 
Consequently, since $F_{r}$ represents the flux, \ie, the sum of photon contributions in all directions, this means that the $F_{r}$ value is zero.
Nevertheless, in the shock wake region located in front of the region where recombinations take place, the radiative flux cannot be in equilibrium because 
the gas is partially optically thin. 
This region includes the precursor, the shock front, and the first part of the wake where the thermalization occurs between heavy particles and electrons. 
In a first approximation, we can consider the wake recombination region as a blackbody at temperature $T$ and producing an equilibrium radiative flux,

\begin{equation}
F_{r}=F^{*}_{r}\equiv\sigma_{r}T^{4}
\label{eq:F*}
,\end{equation}

\noindent in the shock front direction. Because all atoms do not recombine at the same temperature and electronic density, $T$ must be 
considered as an {effective temperature} integrating large gradients of the physical variables present in the recombination region. 
Thus, from these thermodynamical equilibrium relations, we have

\begin{equation}
\frac{F^{*}_{r}/\rho\,u}{e^{*}_{r}}=\frac{1}{3}\frac{F^{*}_{r}/\rho\,u}{p^{*}_{r}/\rho}=\frac{c}{4u}\gg1\,.
\label{eq:F*e*}
\end{equation}

 It should be noted that if the radiation field is larger in one direction (that of the outward flow), then $p_{r}$ will exceed its isotropic limit 
of $1/3e_{r}$ (see, e.g., \cite{miha84} \S66). 
In the extreme limit case occurring in plane waves, and more approximately in the outermost layers of the very extended atmosphere, 
we have $p_{r}=e_{r}$ (see \cite{miha84} \S66). 
Thus, in any case, $p_{r}$ remains close to $e_{r}$. 
Finally, at thermodynamical equilibrium, Eq.\,(\ref{eq:F*e*}) shows that the radiative flux is always the main radiative contribution to the energy 
Rankine-Hugoniot relation.

This is explained by the fact that the speed of light is usually much higher than the relative post-shock velocity $u\equiv\,V_{\textrm{shock}}-{\textrm{v}}$. 
This is almost always  the case in  stellar atmospheres, where $u$ rarely exceeds 300\kms (0.1\% of $c$). 
We note that for extremely fast shocks, such as  supernovae explosions, where the initial shock front velocity is around 5\% of the speed of light, the 
radiative pressure and energy are nearly equal to the gas pressure and energy. 
Thus, it is only for very fast flows of this kind that the three radiative terms must be considered.

Consequently, hereafter the only radiative term that we consider in the total energy Rankine-Hugoniot relation is the radiative flux $F_{r}$: 

\begin{equation}
\frac{1}{2}u^{2}+e_{g}+\frac{p_{g}}{\rho}-\frac{F_{r}}{\rho u}={\rm constant}
\label{eq:rherFr}
\end{equation}

\noindent The gas energy $e_{g}$ is the sum of the energy $e_{t}$ involved into the {translational} degrees of freedom of atoms and the energy $e_{i}$ stocked into their {internal} degrees of freedom 
(excitation, dissociation, ionization):

\begin{equation}
e_{g}=e_{t}+e_{i}\,.
\label{eq:edic}
\end{equation}

\noindent In an ideal gas, the energy $e_{t}$ in translational degrees of freedom is given by

\begin{equation}
e_{t}=\frac{1}{\gamma-1}\frac{p}{\rho}\,,
\label{eq:et}
\end{equation}

\noindent where $\gamma$ is the ratio of specific heat.

If $E$, $D$, and $I$ respectively denote the excitation, dissociation, and ionization specific energies stored in the degrees of freedom of atoms, the internal gas energy $e_{i}$ is

\begin{equation}
e_{i}=E+D+I\,.
\label{eq:edi}
\end{equation} 

If indexes $1$ and $2$ denote the gas just before and just after the shock front, and if we neglect the gas energy in the unperturbed gas, then we have

\begin{equation}
\frac{1}{2}u_1^{2}-\frac{F_{r1}}{\rho_1\,u_1}=\frac{\gamma}{\gamma-1}\frac{p_2}{\rho_2}+e_{i2}.
\label{eq:rh3a}
\end{equation} 

If we consider the medium 2, the shock cools very quickly, and then we have $T_2 = T_1$. 
In this case (isothermal shock), all the energy stored in the internal degrees of freedom of atoms is totally dissipated after the shock front. 
In addition, if the ballistic movement of the atmosphere has already been completed before the arrival of the next shock, then the velocity of the gas 
falling down on the star at the level of the shock is zero ($u_1 = V_{\textrm{shock}}$).
In these conditions, the above Rankine-Hugogniot relation reduces to

\begin{equation}
\frac{1}{2}V_{\textrm{shock}}^{2}=\frac{F_{r1}}{\rho_1\,V_{\textrm{shock}}}.
\label{eq:rh3b}
\end{equation} 

This means that the shock cannot radiate more radiative energy flux than

\begin{equation}
F_{r1}\leq \frac{1}{2}{\rho_{1} V_{\textrm{shock}}^{3}}\,.
\label{eq:rh3c}
\end{equation}

\end{appendix}

\end{document}